       \documentstyle[]{l-aa}
       \input psfig.tex

        \newcommand{\Hbeta}{\mbox{${\rm  H_{\beta}}$}}
        \newcommand{\MgFe}{\mbox{${\rm  [MgFe]} $}}
        \newcommand{\MFe}{\mbox{${\rm \langle  Fe \rangle} $}}

	\newcommand{\Msun}{\mbox{${\rm M_{\odot}}$}}
	\newcommand{\Zsun}{\mbox{${\rm Z_{\odot}}$}}

\def\M12{${\rm M_{L,12}} $}
\def\Mg2{${\rm  Mg_{2}} $}

\def\smallskip{\vskip 8pt}
\def\littleskip{\vskip 6pt}

\begin{document}

\thesaurus{}

\title{Spectro-Photometric Evolution of Elliptical Galaxies. III. }

\subtitle{Infall models with gradients in mass density and star formation }

\author {R. Tantalo$^1$, C. Chiosi$^{2,1}$, A. Bressan$^3$, P. Marigo$^1$,
         L. Portinari$^1$}

\institute {
$^1$ Department of Astronomy, Vicolo dell'Osservatorio 5, 35122 Padua,
   Italy\\
$^2$ European Southern Observatory, Karl-Schwarzschild-strasse 2, 
      D-85748, Garching bei Muenchen, Germany\\
$^3$ Astronomical Observatory, Vicolo dell'Osservatorio 5, 35122 Padua, 
   Italy}

\offprints{C. Chiosi }

\date {Received: October 1997, Accepted: }

\maketitle

\markboth{Elliptical Galaxies: Infall models with 
gradients in mass density and star formation }{}

\begin{abstract}
In this study we present a simple  model of elliptical galaxies aimed at 
interpreting the gradients in colours and narrow band indices observed across
these systems. Salient features of the model are the gradients in mass density
and star formation and infall of primordial gas aimed at simulating the
collapse of a galaxy into the potential well of dark matter. Adopting a
multi-zone model we follow in detail the history of star formation, gas
consumption, and chemical enrichment of the galaxy and also allow for the
occurrence of galactic winds according to the classical supernova (and stellar
winds) energy deposit. The outline of the model, the time scale of gas
accretion and rate of star formation as a function of the galacto-centric
distance  in particular, seek to closely mimic the results from Tree-SPH
dynamical models. Although some specific ingredients of the model can be
questioned from many points of view (of which we are well aware),  the model
 has to be considered as a gross tool for exploring the
consequences of different recipes of  gas accretion and star  formation in
which the simple one-zone scheme is abandoned. With the aid of this model we
discuss the observational data on the gradients in metallicity, colours, and
narrow band indices  across elliptical galaxies.


\end{abstract}

\section{Introduction}

Gradients in  broad-band colours and line strength indices have been observed
in elliptical galaxies (cf. Worthey et al. 1992; Gonz\'ales 1993; Davies et al.
1993; Carollo et al. 1993;  Carollo \& Danziger 1994a,b; Balcells \& Peletier 
1994;  Fisher et al. 1995, 1996). Since variations in colours and line 
strength indices are eventually reduced to variations in age and chemical 
composition (metallicity), or both, of the underlying stellar populations, the 
interpretation of the gradients bears very much on the general mechanism of 
galaxy formation and evolution. Unfortunately, separating age from metallicity 
effects is a cumbersome affair, otherwise known as the {\it age-metallicity 
degeneracy} (cf. Worthey 1994 and references therein) which makes it difficult 
to trace back the history of star formation and chemical enrichment both in 
time and space. Despite this intrinsic difficulty, line strength indices such 
as $\rm H_{\beta}$, $\rm Mg_2$, 
$\rm \langle Fe \rangle$, and \MgFe\ and broad-band 
colours and their gradients are customarily used to infer age and composition 
and their variations across galaxies.

Particularly significant in this context, is the different slope of the \Mg2\ 
and $\rm \langle Fe \rangle$ gradients observed across elliptical galaxies. The 
gradient in \Mg2\ is often steeper than the gradient in 
$\rm \langle Fe \rangle$, 
which is customarily interpreted as indicating that the ratio [Mg/Fe] is 
stronger toward the center. Similar conclusion is reached interpreting the 
systematic increase of \Mg2\ with the galaxy luminosity (mass): the ratio 
[Mg/Fe] seems  to increase with the galaxy mass (the so-called 
$\alpha$-enhancement). This observational hint 
has been taken as one of the most important 
constraints to be met by any chemo-spectro-photometric model of elliptical 
galaxies (cf. Matteucci 1994, 1997 for recent reviews of the subject). 

Owing to the primary importance of this topic, Tantalo et al. (1998) addressed
the question to which extent the gradients in \Mg2\ and 
$\rm \langle Fe \rangle$ 
translate into gradients in chemical abundances and abundance ratios. To this 
aim, the above indices were calculated for a mix of stellar populations with 
known pattern of abundances as a function of the age and position to check 
whether a higher [Mg/Fe] finds one-to-one correspondence with a stronger \Mg2\ 
as compared to $\rm \langle Fe \rangle$. 

The need of a simple tool to follow the chemical history of a galaxy both in 
time and space spurred the model presented in this study. The bottom line is to
abandon the widely adopted one-zone approximation however without embarking in 
a fully dynamical description which would hamper the quick analysis of the 
problem. The model allows for the infall of primordial gas into the potential 
well of dark matter (seeking to closely mimics results from fully 
hydrodynamical models) and the existence of gradients in mass density and star 
formation whose net result is to given rise to gradients in age and composition
of the underlying stellar populations.

The plan of the paper is as follows. Section 2 sketches the model and presents
the basic notation. Section 3 describes the spatial distribution of luminous
and dark matter and their gravitational potentials. Section 4 presents in some 
detail the equations governing the chemical evolution together with the law of 
star formation and the initial mass function we have adopted. Section 5 deals 
with our modelling of the collapse and derives the law for the gas accretion 
time scale, and the specific efficiency of star formation as a function of the 
galacto-centric distance. Section 6 presents the empirical mass-radius 
(effective and total) relationships we have derived from fitting observational 
data. Section 7 clarifies some details of the mass zoning of the models. 
Section 8 deals with galactic winds and summarizes our prescription for the 
energy injection by supernova explosions, and stellar winds. Section 9 presents
the general properties of the models and examines the internal consistency of 
the results. Specifically, it shows the evolution of the gas content, star 
formation rate and metallicity, and the spatial gradients in metallicity and 
relative distribution of stars per metallicity bin. Section 10 contains the 
photometric properties of the models (broad band colours and line strength 
indices), i.e. the color-magnitude relation, the mass to blue luminosity ratio,
the UV excess, the surface brightness profiles, the gradients in broad-band 
colours and indices $\rm Mg_2$ and 
$\rm \langle Fe \rangle$, and the plane \Hbeta-\MgFe\
putting into evidence some difficulties encountered with the gradients in these
quantities. Finally, Section 11 draws some concluding remarks.

%%%%%%%%%%%%%Figure 1
\begin{figure}
%\picplace{9cm}
%\psfig{file=f_dark_lum.ps,height=9.0truecm,width=8.5truecm}
\psfig{file=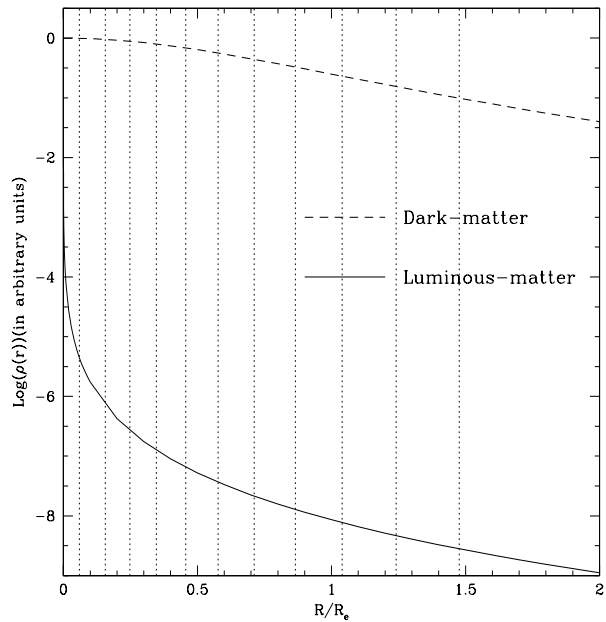,height=9.0truecm,width=8.5truecm}
\caption{The density profiles of baryonic and dark material in the prototype 
galaxy of 1$\times M_{L,T,12}$. For the baryonic component the asymptotic 
density is displayed (see the text for more details)}
\label{f_sketch}
\end{figure}

%%%%%%%%%%%%%%%%%%%%%%%%%%%%%%%%%%%%%%%%%%%%%%%%%
\section{ Modeling elliptical galaxies }

%****************************************************
\subsection{Sketch of the models and basic notation}
Elliptical galaxies are assumed to be made of baryonic and dark material both 
with spherical distributions but different density profiles. Let $M_{L,T}(T_G)$
and $M_{D,T}(T_G)$ be the total luminous and dark mass, respectively, existing 
in the galaxy at the present time ($T_G$ is the galaxy age). The two components
have different effective (half mass) radii, named $R_{L,e}(T_G)$ and 
$R_{D,e}(T_G)$ (thereinafter shortly indicated as $R_{L,e}$ and $R_{D,e}$), and
their masses are in the ratio $M_{D,T}(T_G)/M_{L,T}(T_G)=\theta$. Although 
$\theta$ may vary from galaxy to galaxy, for the purposes of this study it is 
supposed to be constant.

An essential feature of the model is that while dark matter is assumed to have
remained constant in time, the luminous material is let  fall 
at a suitable rate (to be defined below) into the potential well of the former.
Owing to this hypothesis, no use is made of dark matter but for the calculation
of the gravitational potential and the whole formulation of the problem stands 
on the mass and density of luminous material which are let grow with time from 
zero to their present day value.
\littleskip

\noindent
{\it The asymptotic model}. The model whose radial density profile upon 
integration
over radius and time yields the mass $M_{L,T}(T_G)$ is referred to as the 
asymptotic model. If $\rho_{L}(R,t)$ is the radial density profile of luminous 
matter at any age $t$ and $\dot \rho_{L}(R,t)$ is the rate of variation by gas 
accretion, the following relation holds  

%%%%%%%%%%Equation 1
\begin{equation}
M_{L,T}(T_G) = \int _0^{T_G} dt~\int_0^{R_{L,T_G}} 4 \pi ~R^2~
                         \dot\rho_L(R,t) ~dR  .  
\label{mass_tot}
\end{equation}

\noindent
{\it Mass and space zoning}. The asymptotic model is divided into a number of 
spherical shells with equal value of the asymptotic luminous mass, typically 
5\% of $M_{L,T}(T_G)$. Since the density $\rho_{L}(R,T_G)$ is changing with 
radius (decreasing outward), the thickness and volume of the shells are not the
same. They are indicated by

\begin{displaymath}
\Delta R_{j/2} =  R_{j+1} - R_{j} 
\end{displaymath} 
\begin{displaymath}
\Delta V(R_{j/2}) = \frac{4}{3} \pi ({R}_{j+1} ^{3} - {R}_{j} ^{3} )
\label{rag_vol}
\end{displaymath}

\noindent
where $R_{j+1}$ and $R_{j}$ are the outer and inner radii of the shells, and 
$j=0,..J-1$ with $R_0=0$ (the center) and $R_J=R_{L,T_G}$ (the total radius).
The radii $R_j$ are not yet defined. 

Thereinafter we will make use of the following notation and change of the 
radial coordinate:

\begin{itemize}

\item {Each zone of a model is identified by its mid radius 
$R_{j+1/2}=(R_{j+1}+R_{j})*0.5$ shortly indicated by $R_{j/2}$.}
\littleskip

\item {Radial distances are expressed in units of the effective radius of 
the luminous material in the asymptotic model, i.e $r_{j/2}=R_{j/2}/R_{L,e}$.}
\littleskip

\item {All masses are expressed in units of $10^{12} \times M_{\odot}$. Finally
galactic models are labelled by their asymptotic total luminous mass 
$M_{L,T}(T_G)$ in the same units, shortly indicated by $M_{L,T,12}$.}
\end{itemize}

Let us now define for each shell the mean density of total luminous material 
$\overline{\rho}_{L}(r_{j/2},t)$, of stars $\overline{\rho}_{s}(r_{j/2},t)$,
and gas $\overline{\rho}_{g}(r_{j/2},t)$, so that the corresponding masses are

\begin{displaymath}
 ~~~~~~~~~~\Delta M_L(r_{j/2},t) =  {\overline \rho}_{L}(r_{j/2},t) \times 
                 \Delta V(r_{j/2}) \times R_{L,e}^3 
\end{displaymath}
\begin{displaymath}
 ~~~~~~~~~~\Delta M_g(r_{j/2},t) =  {\overline \rho}_{g}(r_{j/2},t) \times 
                 \Delta V(r_{j/2})  \times R_{L,e}^3 
\end{displaymath}
\begin{displaymath}
 ~~~~~~~~~~\Delta M_s(r_{j/2},t) =  {\overline \rho}_{s}(r_{j/2},t) \times 
                \Delta V(r_{j/2})  \times R_{L,e}^3 . 
\label{mass_shell}
\end{displaymath}

\noindent
By definition 
%%%%%%%%%%%Equation 3
\begin{equation}
\Sigma_{j=0} ^{J-1} \Delta M_L(r_{j/2}, T_G)= M_{L,T}(T_G)
\label{sum_mass}
\end{equation}
and 
%%%%%%%%%%%Equation 4
\begin{equation}
\Delta M_g(r_{j/2},t) + \Delta M_s(r_{j/2},t) = \Delta M_L(r_{j/2},t).
\end{equation}

Identical relationships can be defined for the dark matter by substituting
its constant density profile. Since there would be no direct use of these
relations, we just say that the space zoning of the dark
matter distribution is the same as for the luminous component, so that
the contribution  of dark matter to the total gravitational potential in 
each zone is properly calculated (see below).
\littleskip

\noindent
{\it The dimension-less formulation}. Finally, we define the dimension-less variables 

%%%%%%%%%%%%Equation 5
\begin{displaymath}
G_g(r_{j/2},t) = {\Delta M_g(r_{j/2},t) \over \Delta M_L(r_{j/2},T_G) } =  
   { { \overline \rho}_{g}(r_{j/2},t)  \over  
               {\overline \rho}_L(r_{j/2},T_G)  } 
\end{displaymath}
\begin{equation}
G_s(r_{j/2},t) = {\Delta M_s(r_{j/2},t) \over \Delta M_L(r_{j/2},T_G) } =  
   { { \overline \rho}_{s}(r_{j/2},t)  \over  
                {\overline \rho}_L(r_{j/2},T_G)  } 
\label{mass_g}
\end{equation}

\noindent
where ${\overline \rho}_L(r_{j/2},T_G)$ is the mean density of luminous mass
within each shell at the present time.

\noindent
Furthermore for each shell we introduce the gas components 
$G_{g,i}(r_{j/2},t)= G_g(r_{j/2},t)\times X_i(r_{j/2},t)$ where 
$X_i(r_{j/2},t)$ are the abundances by mass of the elemental species $i$. 
Summation of  $X_i(r_{j/2},t)$ over all the species in each shell is equal to 
unity.

%***************************************
\subsection{The Infall scheme}
The density of the luminous component in each shell 
is let increase with time according to

%%%%%%%%%%%%%Equation 6
\begin{equation}
\left[ \frac{d {\overline{\rho}}_{L} (r_{j/2},t)}{dt} \right] = 
        {\overline{\rho}}_{L0}(r_{j/2}) exp\left[-\frac{t}
                {\tau(r_{j/2})}\right]
\label{infall}
\end{equation}

\noindent
where $\tau(r_{j/2})$ is the local time scale of gas accretion for which a 
suitable prescription is required.

The function ${\overline{\rho}}_{L0}(r_{j/2})$ is fixed by imposing that at the
present galactic age $T_{G}$ the density of luminous material in each shell 
has grown to the value given by the adopted profile 
$\overline{\rho}_{L}(r,T_{G})$ 

%%%%%%%%%%%%%Equation 7
\begin{equation}
{\overline{\rho}}_{L0}(r_{j/2}) = \frac{\overline{\rho}_{L}(r_{j/2},T_{G})}
           {\tau(r_{j/2}) [1 -exp(-\frac{T_{G}}{\tau(r_{j/2})})]} 
\label{rho_zero}
\end{equation}

\noindent
It follows that the time dependence for $\overline{\rho}_{L}(r_{j/2},t)$ is 
given by

%%%%%%%%%%%%%Equation 8
\begin{equation}
\overline{\rho}_{L}(r_{j/2},t) = 
\frac{\overline{\rho}_{L}(r_{j/2},T_{G})}{[1 -
     exp(-\frac{T_{G}}{\tau(r_{j/2})})]} \times 
     \left[1 - exp(-\frac{t}{\tau(r_{j/2})})\right]
\label{dens}
\end{equation}

For the asymptotic mass density in each shell we adopt the geometric mean of 
the values at inner and outer radii, i.e. 

%%%%%%%%%%%%%Equation 9
\begin{equation}
\overline{\rho}_{L}(r_{j/2},T_G) = \sqrt{\rho_{L}(r_{j+1},T_G) \times
\rho_{L}(r_{j},T_G)}.
\label{rho_squa}
\end{equation}

\noindent
To summarize, each shell is characterized by:
\begin{itemize}

\item The radius $r_{j/2}$.
\littleskip

\item The asymptotic mass $\Delta M_{L}(r_{j/2},T_{G})$, which is a suitable
fraction of the total asymptotic luminous mass $M_{L,T,12}$.
\littleskip

\item The mass of dark matter $\Delta M_{D}(r_{j/2},T_{G})$. Since this mass 
is constant with time no other specification is required.
\littleskip

\item The asymptotic mean  density $\overline{\rho}_{L}(r_{j/2},T_{G})$ of
baryonic mass (gas and stars). 
\littleskip

\item The gravitational potential for the luminous component 
$\varphi_{L}(r_{j/2},t)$ varying with time, and the corresponding gravitational
potential of dark-matter $\varphi_{D}(r_{j/2},T_{G})$, constant with time. Both
will be defined below.
\end{itemize}

%%%%%%%%%%%%%%%%%%%%%%%%%%%%%%%%%%%%%%%%%%%%%%%%%%%%%%%%%%%%%%%%%%%%%%%%
\section{The spatial distribution of luminous and dark matter, and 
gravitational binding energies }

{\it Density profile of the luminous matter}. The asymptotic spatial distribution of 
luminous matter is supposed to follow the Young (1976) density profile. This 
is derived from assuming that the $R^{1/4}$-law holds and the mass to 
luminosity ratio is constant throughout the galaxy (cf. Poveda et al. 1960, 
Young 1976, Ciotti 1991). We remind the reader that the density 
$\rho_{L}(R,T_G)$ and the gravitational potential $\varphi_L(R,T_G)$ are
expressed by Young (1976) as a function of the effective radius for which a 
suitable relationship with the total luminous mass is required (see below).
 
The adoption of the Young (1976) density profile imposes that the resulting
model at the age $T_G$ has (i) a radially constant mass to luminosity ratio; 
(ii) a luminosity profile obeying the $R^{1/4}$ law. 
\littleskip

\noindent 
{\it Density profile of the dark matter}. The mass distribution and 
gravitational 
potential of the dark-matter are derived from the density profiles by Bertin 
et al. (1992) and Saglia et al. (1992) however adapted to the Young formalism 
for the sake of internal consistency. In brief we start from the density law

%%%%%%%Equation 10
\begin{equation}
\rho_{D} (R) = \frac{\rho_{D,0} \times
                 R_{D,0}^{4}}{(R_{D,0}^{2} + R^{2})^{2}}
\label{rho_dark}
\end{equation}
\noindent
where $R_{D,0}$ and $\rho_{D,0}$ are two scale factors of the distribution.
The density scale factor $\rho_{D,0}$ is derived from imposing the relation  
$M_{D,T} = \theta M_{L,T}$ and the definition of  $M_{D,T}$ by means of its
density law

%%%%%%%%%%%Equation 11
\begin{equation}
M_{D,T} = 4\pi \int_{0}^{\infty} R^{2}\rho(R) dR = 
4\pi \rho_{D,0} R_{D,0}^{3} m(\infty)
\label{ddm}
\end{equation}

\noindent
with 
%%%%%%%%%%Equation 12
\begin{equation}
m(\infty) = \int_{0}^{\infty}
\frac{R^{2}}{r_{D,0}^{3}\left(1+\left(\frac{R}{R_{D,0}}
                            \right)^{2}\right)^{2}}dR
\label{mass_infty}
\end{equation}

\noindent
This integral is solved numerically. Finally, the density profile of 
dark-matter is 

%%%%%%%%%%Equation 13
\begin{equation}
\rho_{D}(R) = \frac{M_{D,T}}{m(\infty)}\frac{1}{4\pi R_{D,0}^{3}}
\frac{1}{\left(1+\left(\frac{R}{R_{D,0}}\right)^{2}\right)^{2}} .
\label{rho_dark_2}
\end{equation}

The radial dependence of the gravitational potential of dark matter is 

%%%%%%%%%%Equation 14
\begin{equation}
\varphi_{D}(R) = - G \int_{0}^{R} \frac{M_{D}(R)}{r^{2}} dR
\label{poten_dark}
\end{equation}

\noindent
which upon integration becomes

%%%%%%%%%%%%Equation 15
\begin{equation}
\varphi_{D}(R) = - 4\pi G \rho_{D,0} R_{D,0}^{2}
\widetilde{\varphi_{D}}\left(\frac{R}{R_{D,0}}\right)
\label{poten_dark_1}
\end{equation}

\noindent
where $\widetilde{\varphi_{D}}\left(\frac{R}{R_{0}}\right)$ is given by

%%%%%%%%%%%Equation 16
\begin{equation}
\int_{0}^{R/R_{D,0}}\frac{m(R/R_{D,0})}
             {R_{D,0}\left(\frac{R}{R_{D,0}}\right)^{2}}dR
\label{poten_dark_2}
\end{equation}
This integral is solved numerically and stored as a look-up table function of 
$R/R_{D,0}$. 
\littleskip

We assume $R_{D,0}= \frac{1}{2} R_{D,e}$, where $R_{D,e}$ is the effective
radius of dark matter. This can be derived from relation(\ref{ddm}) looking
for the radial distance within which half of the dark-matter mass is contained.
Finally, all the models below are calculated adopting $\theta= 5$ in the ratio 
$M_{D,T}(T_G)/M_{L,T}(T_G)=\theta$.
\littleskip

\noindent
{\it The gravitational binding energies}. The binding gravitational energy for the gas 
in each shell is given by:

%%%%%%%%%%%%%Equation 17
\begin{displaymath}
\Omega_{g}(r_{j/2},t)= \overline{\rho}_{g}(r_{j/2},t) 
\Delta V(r_{j/2}) \varphi_{L}(r_{j/2},t) +
\end{displaymath}
\begin{equation}
~~~~~~~~~~~~~~~~~~~\overline{\rho_{g}}(r_{j/2},t) \Delta V(r_{j/2}) 
                   \varphi_{D}(r_{j/2},T_{G})
\label{gas_pot}
\end{equation}

%%%%%%%%%%%%%%%%%%%%%%%%%%%%%%%%%%%%%%%%%%%%%%%
\section{The chemical equations}

The chemical evolution of elemental species is governed by the same set of
equations as in Tantalo et al. (1996, TCBF96) and Portinari et al. (1998)
however adapted to the density formalism and improved as far the ejecta and 
the contribution from the Type Ia and Type II supernovae are concerned 
(cf. Portinari et al. 1998). Specifically, we follow in detail the evolution 
of the abundance of thirteen chemical elements ( $\rm H$, $\rm ^{4}He$, 
$\rm ^{12}C$, 
$\rm ^{13}C$, $\rm ^{14}N$, $\rm ^{16}O$, $\rm ^{20}Ne$, 
$\rm ^{24}Mg$, $\rm ^{28}Si$, $\rm ^{32}S$, 
$\rm ^{40}Ca$, $\rm ^{56}Fe$, and the isotopic neutron-rich elements $nr$ obtained by 
$\alpha$-capture on $\rm ^{14}N$, specifically $\rm ^{18}O$, 
$\rm ^{22}Ne$, $\rm ^{25}Mg$). 
Furthermore, the stellar yields in usage here take  into account the effects of
different initial chemical compositions (cf. Portinari et al. 1998). 

The equations governing the time variation of the $G_{i}(r,t)$ and hence
$X_{i}(r,t)$ are:

%%%%%%%%%%%%%Equation 18
\begin{displaymath}
\frac{dG_{i}(r_{j/2},t)}{dt}= - X_{i}(r_{j/2},t) \Psi(r_{j/2},t) + 
\end{displaymath}

\begin{displaymath}
 \int_{M_{min}}^{M_{Bm}}\Psi(r_{j/2}, t-\tau_M)Q_{M,i}(t-\tau_M)\phi(M)dM +
\end{displaymath}

\begin{displaymath}
\Lambda \int_{M_{Bm}}^{M_{BM} } \phi(M_B) dM_B \times 
\end{displaymath}

\begin{displaymath}
[\int_{\mu_{min}}^{0.5} f(\mu) 
            \Psi(r_{j/2},t-\tau_{M_2})Q_{M,i}(t-\tau_{M_2})d\mu ] +
\end{displaymath}

\begin{displaymath}
(1-\Lambda) \int_{M_{Bm} }^{M_{BM}} \Psi(r_{j/2}, t-\tau_M) Q_{M,i}(t-\tau_M) 
         \phi(M) dM +
\end{displaymath}

\begin{displaymath}
\int_{M_{BM} }^{M_{max}} 
\Psi(r_{j/2},t-\tau_M) Q_{M,i}(t-\tau_M) \phi(M) dM  +
\end{displaymath}

\begin{equation}
 \left[ \frac{dG_{i}(r_{j/2},t)}{dt}
\right]_{inf}
\label{eq_chem}
\end{equation}

\noindent
where all the symbols have their usual meaning. Specifically $\Psi(r_{j/2},t)$ 
is the normalized rate of star formation for the shell, $Q_{M,i}(t)$ are the 
restitution fractions of the elements $i$ from stars of mass $M$ (cf. Talbot 
\& Arnett 1973), $\phi(M)$ is the initila mass function (IMF), 
whose lower and upper mass limits are 
$M_{min}$ and $M_{max}$ (see below). $\tau_M $ is the lifetime of a star of 
mass $M$, for which the dependence on the initial chemical composition is also 
taken into account using the tabulations by Bertelli et al. (1994). The 
various integrals appearing in eq.(\ref{eq_chem}) represent the separated 
contributions of Type II and Type Ia supernovae as introduced by Matteucci \& 
Greggio (1986). In particular, the second integral stands for all binary 
systems having the right properties to become Type Ia supernovae. $M_B$$_{m}$ 
and $M_B$$_{M}$ are the lower and upper mass limit for the total mass $M_B$ of 
the binary system, $f(\mu)$ is the distribution function of their mass ratios, 
and $\mu_{min}$ is the minimum value of this, finally $\Lambda$ is the fraction
of such systems with respect to the total. It is assumed here that binary 
stars as a whole obey the same IMF of single stars. We adopt 
$B_{m}=3 M_{\odot}$, $B_{M}=12 M_{\odot}$, and $\Lambda=0.02$. The stellar 
ejecta are from Marigo et al. (1996, 1997), and Portinari et al. (1998) to 
whom we refer for all details.

%**********************************************
\subsection{Star formation rate and IMF}

The stellar birth rate, i.e. the number of stars with mass $M$ born in the
interval $dt$ and mass interval $dM$, is expressed by 

%%%%%%%%%%%%%Equation 21
\begin{equation}
dN =\Psi(R,t,Z) \phi(M)~dM~dt  
\label{birth}
\end{equation}

\noindent
with  obvious meaning of the symbols.

Neglecting the possible dependence on the gas composition,  
the rate of star formation (SFR), $ \Psi(r_{j/2},t,Z)$, 
 is assumed to depend on the gas density
according to the Schmidt (1959) law (see also Larson 1991) 

%%%%%%%%%%%Equation 19
\begin{equation}
\Psi(r_{j/2},t) =  {d {\overline\rho}_g(r_{j/2},t) \over dt }=\nu(r_{j/2}) 
                             {\overline\rho}_g(r_{j/2},t)^{\kappa}
\label{drho_dt}
\end{equation}

\noindent
where  the specific efficiency of star formation 
$\nu(r_{j/2})$ is a suitable function to be specified below. 

\noindent
Upon normalization, the star formation rate becomes:
%%%%%%%%%%%%Equation 20
\begin{equation}
\Psi(r_{j/2},t)= \nu(r_{j/2},t) [{ {\overline\rho}_{L}(r_{j/2},T_G) }]^{k-1}
  G_g(r_{j/2},t)^{k} .
\label{sfr_nor}
\end{equation}

\noindent
All the models we are going to describe are for $\kappa=1$.

For the IMF  $\phi(M)$ we  have adopted the Salpeter law:

%%%%%%%%%%%%%Equation 22
\begin{equation}
{\rm \phi(M) \propto  M^{-x}  }  
\label{imf}
\end{equation}

\noindent
where $x~=~2.35$. The IMF is normalized by imposing the fraction $\zeta$ of 
mass in the IMF above a certain value $M_*$. i.e.  

%%%%%%%%%%%%% equation 23
\begin{equation}
\zeta =  \frac{\int_{M_*}^{M_U}\phi(M){\times}M{\times}dM}
{\int_{M_L}^{M_U}\phi(M){\times}M{\times}dM}
\label{zeta}
\end{equation}

\noindent
A useful choice for $M_*$ is the minimum star mass contributing to the 
nucleo-synthetic enrichment of the interstellar medium over a time scale 
comparable to the total lifetime of a galaxy. This is approximately equal to 
$1$\Msun. The upper limit is $M_{U}=120 M_{\odot}$ corresponding to the maximum
mass in our data base of stellar models. The parameter $\zeta$ is fixed by 
imposing that our models match the mean mass to blue luminosity ratio for 
elliptical galaxies (Bender et al. 1992, 1993), this yields $\zeta=0.50$. With 
this normalization the minimum star mass of the IMF is $M_{L} \sim 0.18$\Msun.

%%%%%%%%%%%%%%%%%%%%%%%%%%%%%%%%%%%%%%%%%
\section{Modelling the collapse}
To proceed further, we need to supply our models with the radial dependence of
the time scale of gas accretion $\tau(r_{j/2})$ and specific efficiency of
star formation $\nu(r_{j/2})$. 

\begin{figure}
%\picplace{9cm}
%\psfig{file=f_ra_ve.ps,height=9.0truecm,width=8.5truecm}
\psfig{file=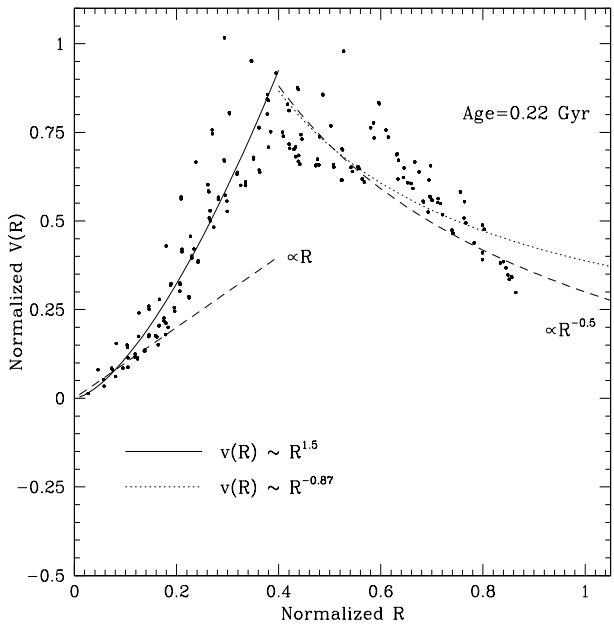,height=9.0truecm,width=8.5truecm}
\caption{The radial velocity $v(R/R_0)$ versus radius $R/R_0$ 
relationship for a 
Tree-SPH model of the adiabatic collapse of a galaxy with total mass (baryonic 
and dark matter) of $10^{12} M_{\odot}$ from Carraro et al.\ (1997). The 
velocity and radius are normalized to $v^*$ and $R_0$ as described in the text. 
The full dots are the results of the numerical calculations. The {\em solid } 
and {\em dotted } lines are the best fits of the data: 
$v(R) \propto R^{1.5}$ for
the inner core and $v(R) \propto  R^{-0.87}$ for the external regions. The 
{\em dashed lines} are the same but for the strict homologous collapse and 
free-fall description}
\label{f_ra_ve}
\end{figure}

The dynamical models of galaxy formation and evolution with the Tree-SPH 
technique (cf. Carraro et al. 1997 and references therein) hint the solution to
this problem. In brief, looking at the paradigmatic case of the adiabatic 
collapse of a galaxy (dark plus baryonic material) of $10^{12} M_{\odot}$,
initial mean density of $1.6 \times 10^{-25}$ g cm$^{-3}$, free-fall time scale
of 0.25 Gyr, and  age of 0.22 Gyr, we notice that the radial velocity $v(R)$ as
a function of the radial distance $R$ starts from zero at the center, increases
to a maximum at a certain  distance, and then decreases again moving further 
out. The situation is shown in Fig.~\ref{f_ra_ve}, where the velocity is units 
$v^*=\sqrt{[GM/R_0^2]}$), the distance is units of the initial radius 
$R_0=100$ kpc, 
and the maximum occurs at $R/R_0 \simeq 0.4$ (for this particular model).

This reminds the core collapse  in a massive star (cf. Bethe 1990),
which obeys the following scheme

\begin{itemize}

\item homologous collapse in all regions internal to a certain value of the 
radius ($R^{*}$): $v(R) \propto R$;
\littleskip

\item free-fall outside: $v(R) \propto R^{-\frac{1}{2}}$;
\end{itemize}

\noindent
where $R^{*}$ is the radius at which the maximum velocity occurs.

How does the above simple scheme compare with the results of numerical
calculations? To this aim, in Fig.~\ref{f_ra_ve} we plot the best fit of the 
data from the numerical model for the two branches of the velocity curve and 
compare them with the above relationships. In this particular example, the 
slope along the ascending branch ($R/R_0 < 0.4$) is  $1.5\div2$ instead of 1, 
whereas that along the descending branch ($R/R_0 > 0.4$) is --0.87 instead of 
--0.5. 

A close inspection of the numerical Tree-SPH models reveals that neither the
slopes of the velocity branches nor the radius of the peak velocity are 
constant in time. Therefore we will consider all of them as free parameters and
cast the problem in a general fashion suited to our aims.

%%%%%%%%%%%%Figure 2

Let us express the velocity {\it v(R)} as
\begin{displaymath}
       v(R) = c_1 \times R^{\alpha}  ~~~~~~~~~~ {\rm for}~~ R \leq ~~ R^*
\end{displaymath}
\begin{displaymath}
       v(R) = c_2 \times R^{- \beta} ~~~~~~~~ {\rm for}~~ R > ~~ R^*
\end{displaymath}
(where $c_1$, $c_2$, $\alpha$ and $\beta$ are suitable constants), and the 
time scale of accretion as

\begin{displaymath}
\tau(R)  \propto { R \over v(R) } 
\end{displaymath}

Many preliminary models calculated with the above recipe, of which no detail is
given here for the sake of brevity, indicate that $\alpha=2$ and $\beta=0.5$ 
are good choices. The value $\alpha=2$ is indeed taken from the Tree-SPH models
whereas $\beta=0.5$ follows from the core collapse analogy. The determination
of the constants $c_1$ and $c_2$ is not required as long as we seek for scaling
relationships. Therefore the time scale of gas accretion can be written as 
proportional to some arbitrary time scale, modulated by a correction term 
arising from the scaling law for the radial velocity. For the time scale 
base-line we can take the free-fall time scale $t_{ff}$ referred to the whole 
system. Passing to our notation for radial distances we get 

%%%%%%%%%%%%% equation 24
\begin{equation}
\tau(r_{j/2}) = 
     t_{ff} \times \frac{r^{*}}{r_{j/2}} ~~~~~~~~~~~~
           { \rm if} ~~ r_{j/2} \leq r^{*} 
\label{tau_ff_1}
\end{equation}

%%%%%%%%%%%% equation 25
\begin{equation}
\tau(r_{j/2}) = 
     t_{ff} \times (\frac{r_{j/2}}{r^{*}})^{3/2} ~~~~~ 
                 { \rm if }~~ r_{j/2} > r^{*} 
\label{tau_ff_2}
\end{equation}

\noindent
For the free-fall time scale $t_{ff}$ we make use of the relation by Arimoto
\& Yoshii (1987)

%%%%%%%%%%%%%Equation 26
\begin{equation}
t_{ff} = 0.0697 \times  M_{L,T,12}^{0.325} ~~~~~~~~~~~{\rm  Gyr}.
\label{tff}
\end{equation}
\noindent
Finally, we take $r^*={1 \over 2}$ for the sake of simplicity. Other 
choices are obviously possible. They would not change the main qualitative 
results of this study.

%%%%%%%%%Table 1 (tau and nu) %%%%%
\begin{table*}
\vskip 0.3 cm
\begin{center}
\vskip 0.2 cm
\caption{The radial dependence of $\tau(r_{j/2})$ and $\nu(r_{j/2})$ in 
galactic models of different asymptotic luminous mass as indicated. The 
collapse time scales $\tau(r_{j/2})$ are in Gyr. The galactic baryonic masses 
are in units of $10^{12} M_{\odot}$.}
\vskip 0.25 cm
\scriptsize
\begin{tabular*} {145mm} {c l|  c r| c r| c r| c r| c r| c r}
\hline
\hline
 & & & & & & & & & & & & & \\
\multicolumn{2}{c|}{Region} &
\multicolumn{2}{c|}{$3M_{L,T,12}$} &
\multicolumn{2}{c|}{$1M_{L,T,12}$} &
\multicolumn{2}{c|}{$0.5M_{L,T,12}$} &
\multicolumn{2}{c|}{$0.1M_{L,T,12}$} &
\multicolumn{2}{c|}{$0.05M_{L,T,12}$} &
\multicolumn{2}{c}{$0.005M_{L,T, 12}$} \\
 & & & & & & & & & & & & & \\
\hline
 & & & & & & & & & & & & & \\
j & $r_{j+1/2}$ & $\tau$ & $\nu$ & $\tau$ & $\nu$ 
& $\tau$ & $\nu$ & $\tau$ & $\nu$ & $\tau$ & $\nu$  & $\tau$ & $\nu$ \\
 & & & & & & & & & & & & & \\
\hline
 & & & & & & & & & & & & & \\
0& $r_{1/2}$ & 0.74 & 7.1    & 0.52 & 9.0    & 0.42 & 10.4   & 0.25 & 14.7   
& 0.20 & 17.0   & 0.09 & 27.7   \\
1& $r_{3/2}$ & 0.29 & 50.0   & 0.20 & 60.6   & 0.16 & 68.3   & 0.10 & 90.6   
& 0.08 & 102.4  & 0.04 & 154.1  \\
2& $r_{5/2}$ & 0.18 & 111.6  & 0.13 & 132.8  & 0.10 & 148.3  & 0.06 & 191.9  
& 0.05 & 214.6  & 0.03 & 312.3  \\
3& $r_{7/2}$ & 0.13 & 198.6  & 0.09 & 233.3  & 0.07 & 258.5  & 0.04 & 328.7  
& 0.03 & 364.9  & 0.03 & 518.5  \\
4& $r_{9/2}$ & 0.10 & 325.5  & 0.07 & 378.3  & 0.06 & 416.3  & 0.03 & 521.4  
& 0.03 & 575.2  & 0.03 & 800.7  \\
5& $r_{11/2}$& 0.14 & 501.4  & 0.10 & 577.3  & 0.08 & 631.6  & 0.05 & 780.8  
& 0.04 & 856.5  & 0.03 & 1171.1 \\
6& $r_{13/2}$& 0.20 & 753.8  & 0.14 & 860.3  & 0.11 & 936.1  & 0.07 & 1142.8 
& 0.05 & 1247.1 & 0.03 & 1676.6 \\
7& $r_{15/2}$& 0.27 & 1116.0 & 0.19 & 1262.8 & 0.15 & 1366.9 & 0.09 & 1648.8 
& 0.07 & 1790.2 & 0.03 & 2367.9 \\
8& $r_{17/2}$& 0.35 & 1632.9 & 0.24 & 1832.5 & 0.20 & 1973.4 & 0.12 & 2352.9 
& 0.09 & 2542.1 & 0.04 & 3309.8 \\
9& $r_{19/2}$& 0.46 & 2383.7 & 0.32 & 2653.1 & 0.25 & 2842.6 & 0.15 & 3350.2 
& 0.12 & 3602.2 & 0.06 & 4616.9 \\
10&$r_{21/2}$& 0.59 & 3493.2 & 0.41 & 3855.9 & 0.33 & 4110.1 & 0.20 & 4787.8 
& 0.16 & 5122.6 & 0.07 & 6462.5 \\
 & & & & & & & & & & & & & \\
\hline
\hline
\end{tabular*}
\end{center}
\label{tab1}
\normalsize
\end{table*}

%%%%%%%%%%%%%Figure 3
\begin{figure}
%\picplace{9cm}
%\psfig{file=f_tau_ra.ps,height=9.0truecm,width=8.5truecm}
\psfig{file=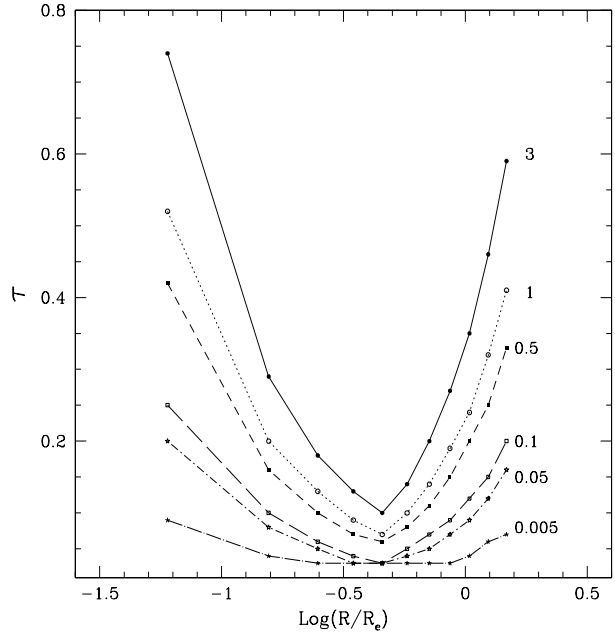,height=9.0truecm,width=8.5truecm}
\caption{The accretion time scale $\tau(r)$ as a function of the 
galacto-centric distance for the models with different asymptotic mass
$M_{L,T,12}$ as indicated}
\label{f_tau_ra}
\end{figure}

%%%%%%%%%%%%%Figure 4
\begin{figure}
%\picplace{9cm}
%\psfig{file=f_nu_ra.ps,height=9.0truecm,width=8.5truecm}
\psfig{file=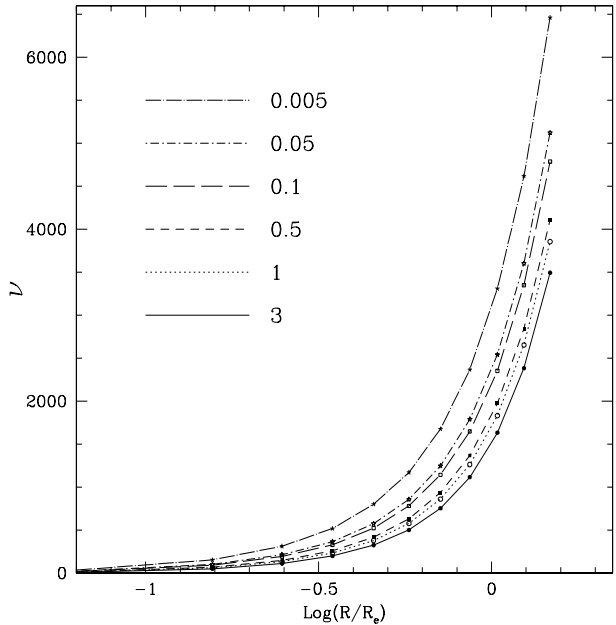,height=9.0truecm,width=8.5truecm}
\caption{The specific efficiency of star formation $\nu(r)$ as a function 
of the galacto-centric distance for the models with different
asymptotic mass $M_{L,T,12}$ as indicated. See the text for more details}
\label{f_nu_ra}
\end{figure}

In order to derive the specific efficiency of star formation $\nu(r_{j/2})$ we 
utilize the simple scale relations developed by Arimoto \& Yoshii (1987) 
however adapted to the density formalism. At the typical galactic densities 
($10^{-22}$ - $10^{-24} $ g~cm$^{-3}$) and considering hydrogen as the dominant
coolant (Silk 1977) the critical Jeans length is much smaller than the galactic
radius, therefore the galaxy gas can be considered as made of many cloud lets 
whose radius is as large as the Jeans scale. If these clouds collapse nearly 
isothermal without suffering from mutual collisions, they will proceed through 
subsequent fragmentation processes till opaque small subunits (stars) will 
eventually be formed. In such a case the stars are formed on the free-fall
time scale. In contrast, if mutual collisions occur, they will be supersonic 
giving origin to layers of highly cooled and compressed material; the Jeans 
scale will then fall below the thickness of the compressed layer and 
fragmentation occurs on the free-fall time scale of the high density layers; 
and finally the whole star forming process is driven by the collision time 
scale. On the basis of these considerations, we take the ratio

%%%%%%%%%%%Equation 29
\begin{equation}
 \sqrt{ \frac{1} {t_{ff} \times t_{col} } } .
\label{nu_prod}
\end{equation}

\noindent
as a measure of the net efficiency of star formation.

Let us express $\nu(r)$ as the product of a suitable yet arbitrary specific 
efficiency $\nu^*$ referred to the whole galaxy times a dimensionless quantity 
$F(r)$ describing as the above ratio varies with the galacto-centric distance. 
An obvious expression for $F(r)$ is the ratio (\ref{nu_prod}) itself normalized
to its central value.

According to Arimoto \& Yoshii (1987) the mean collision time scale referred
to the whole galaxy can be written as

%%%%%%%%%%Equation 30
\begin{equation}
t_{col} = 0.0072  \times M_{L,T,12}^{0.1}   ~~~~~~~~~~~~~~{\rm Gyr}
\label{t_jeans}
\end{equation}

With the aid of this and the relation for the free-fall time scale above we can
calculate $\nu^*$

%%%%%%%%%%%Equation 29
\begin{equation}
\nu^{*} =\left[ \sqrt{ \frac{1} {t_{ff} \times t_{col} } }\right]_{gal} .
\label{nu_star}
\end{equation}

Extending by analogy the above definition of free-fall and collision time 
scales to each individual region, we get

%%%%%%%%%%%%%Equation 27
\begin{equation}
F(r) =  \left( { r_{1/2} \over r_{j/2} } \right) ^{3 \gamma }
          \times 
\left[\frac{ \overline{\rho}_{g}(r_{1/2},T_{G}) }
            {\overline{\rho}_{g}(r_{j/2},T_{G}) }
            \right]^{\gamma} 
\label{nu}
\end{equation}

\noindent
with obvious meaning of the symbols.

In principle, the exponent $\gamma$ could be derived from the mass dependence
of $t_{ff}$ and $t_{col}$, i.e. $\gamma \simeq 0.2$. However, a preliminary 
analysis of the problem has indicated that $F(r)$ must vary with the radial
distance more strongly than this simple expectation. The following relation for
$\gamma$ has been found to give acceptable results as far as gradients in star 
formation, metallicity, colours, etc.. are concerned

%%%%%%%%%%%Equation 28
\begin{equation}
\gamma = 0.98\times (M_{L,12})^{0.02}
\label{alfa_nu}
\end{equation}

\noindent
Finally, the total expression for $\nu(r)$ is 

%%%%%%%%%%%%%Equation 27
\begin{displaymath}
\nu(r) = \left[{ \frac{1} {t_{ff} \times t_{col} } } \right]_{gal}^{0.5}
\times
\end{displaymath}
\begin{equation}
~~~~~~~~~~~~~~~~         \left( { r_{1/2} \over r_{j/2} } \right) ^{3 \gamma }
          \times 
\left[\frac{ \overline{\rho}_{g}(r_{1/2},T_{G}) }
            {\overline{\rho}_{g}(r_{j/2},T_{G}) }
            \right]^{\gamma} ~~{\rm Gyr^{-1}}
\label{nu_tot}
\end{equation}

Table 1 contains the values of $\tau(r_{j/2}$) and  $\nu(r_{j/2})$ as a 
function of the galacto-centric distance for all the galactic models under 
consideration, whereas Figs.~\ref{f_tau_ra} and ~\ref{f_nu_ra} show the same 
in graphical form. As expected, in a galaxy the specific efficiency of star 
formation increases going outward. Likewise, at given relative distance from 
the galactic center, passing from a low to a high mass galaxy.

%%%%%%%%%%%%%%%%%%%%%%%%%%%%%%%%%%%%%%%%%%%%%%%%%%%%%%%%%
\section{The mass-radius relationships}
To proceed further we need to adopt suitable relationships between the 
$R_{L,e}$ and $M_{L,T}$, so that once the the total baryonic mass is assigned, 
the effective radius is derived, and  all the other quantities are properly 
re-scaled.

For the purposes of this study and limited to the case of
$H_{0}=50~{\rm km~sec^{-1}Mpc^{-1} }$, we derive  from the data of Carollo et 
al. (1993), Goudfrooij et al. (1994) the following relation

%%%%%%%%%%%%%%%Equation 31
\begin{equation}
       R_{L,e} = 17.13 \times M_{L,T,12}^{0.557} 
\label{reff_mass}
\end{equation}

\noindent
where $R_{L,e}$ is in kpc.

\noindent
For the same objects and using the diameters from the RC3 catalogue we also 
derived the relation between total radius and mass of the luminous material

%%%%%%%%%%%%%%Equation 32
\begin{equation}
       R_{L,T} = 39.10 \times M_{L,T,12}^{0.402} 
\label{rtot_mass}
\end{equation}

\noindent
in the same units as above.

\begin{figure}
%\picplace{9cm}
%\psfig{file=f_ms_ra.ps,height=9.0truecm,width=8.5truecm}
\psfig{file=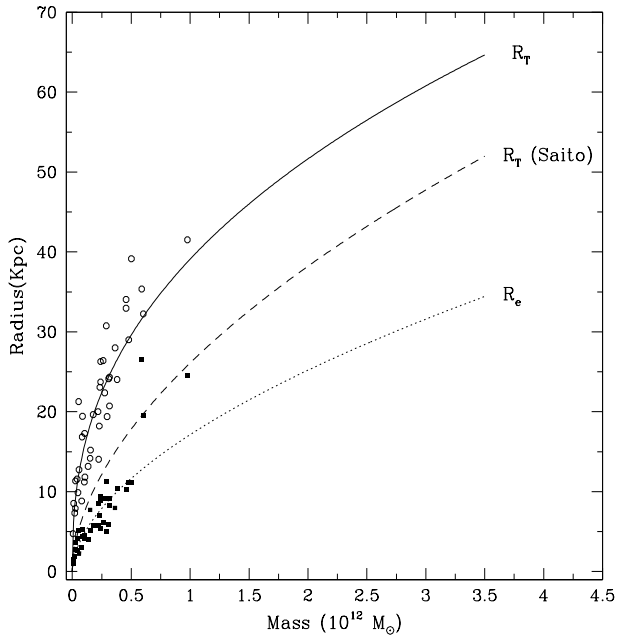,height=9.0truecm,width=8.5truecm}
\caption{The mass-radius relationships derived from the observational data by 
Carollo et al.\ (1993): the {\em open circles} are the total radius, whereas 
{\em filled squares} are the effective radius. The {\em dotted}, and {\em dashed} 
lines show the relationships $M_{L}(R_{T})$ and $M_{L}(R_{e})$. Finally, the 
{\it long-dashed} line displays the relation by Saito (1979a,b) for purposes of 
comparison}
\label{f_ms_ra}
\end{figure}

\noindent
The relations above are displayed in Fig.~\ref{f_ms_ra} and compared with the  
mass radius relation by Saito (1979a,b). Finally, Table 2 lists $R_{L,e}$ and 
$R_{L,T}$ as assigned to each model galaxy.

%%%%%%%%%%%%%Figure 5

%%%%%%%%%Table 2 ( radius of galaxies) %%%%%
\begin{table}
\begin{center}
\vskip 0.2 cm
\caption{Effective and total radii (in kpc) assigned to model galaxies of 
different $M_{L,T,12}$.}
\vskip 0.25 cm
\begin{tabular} {r r r}
\hline
\hline
 & & \\
 $M_{L,T,12}$ & $R_{L,T}$ & $R_{L,e}$ \\
 & & \\
\hline & & \\
 3     &  60.78 & 31.60 \\
 1     &  39.10 & 17.13 \\
 0.5   &  29.60 & 11.64 \\
 0.1   &  15.51 & 4.75  \\
 0.05  &  11.74 & 3.23  \\
 0.005 &  4.66  & 0.90  \\
 & & \\
\hline
\hline
\end{tabular}
\end{center}
\label{tab2}
\end{table}

%%%%%%%%%%%%%%%%%%%%%%%%%%%%%%%%%%%%%%%%%%%%%%%%%%%
\section{Mass zoning of the models} 

The mass zoning of our models is chosen in such a way that within each shell
about 5\% of the luminous mass $M_{L,T,12}$ is contained. From the tabulations
of Young (1976) we derive the corresponding fractionary radius $R/R_{L,e}$,
and from the mass-radius relationships above we fix  the effective radius
$R_{L,e}$, and the real inner and outer radii of each shell. The results are
given in Table 3, whereby the meaning of the various symbols is
self-explanatory.

%%%%%%%%%Table 3   (shell subdivision) %%%%%
\begin{table*}
\vskip 0.3 cm
\begin{center}
\vskip 0.2 cm
\caption{Percentage of luminous mass contained in the sphere of fractionary 
radius $r'/R_{L,e}$, and actual radius $r'$ (in kpc) for model galaxies with 
different $M_{L,T,12}$ as indicated.}
\vskip 0.25 cm
\scriptsize
\begin{tabular*} {100mm} {c| r r r r r r r r}
\hline
\hline
& & & & & & & & \\
 $M_{L,T,12}$ &  & &3        & 1      &0.5       &0.01    &0.05 &  0.005\\
& & & & & & & & \\
\hline
& &  &   &    &    &     & \\
\%$M_{L,T,12}$ & j& $\frac{R}{R_{e}}$ & $R$ & $R$ & $R$ & $R$ & $R$ 
& $R$ \\
 & & & & & & & & \\
\hline
 & & & & & & & & \\
 5   & 0 & 0.1106   & 3.49    & 1.89    & 1.29    & 0.53   & 0.36   & 0.10   \\
 10  & 1 & 0.2005   & 6.33    & 3.43    & 2.33    & 0.95   & 0.65   & 0.18   \\
 15  & 2 & 0.2954   & 9.33    & 5.06    & 0.69    & 1.40   & 0.95   & 0.26   \\
 20  & 3 & 0.3983   & 12.58   & 6.82    & 4.64    & 1.89   & 1.29   & 0.36   \\
 25  & 4 & 0.5127   & 16.20   & 8.78    & 5.97    & 2.44   & 1.66   & 0.46   \\
 30  & 5 & 0.6405   & 20.24   & 11.05   & 7.46    & 3.04   & 2.07   & 0.57   \\
 35  & 6 & 0.7833   & 24.75   & 13.42   & 9.12    & 3.72   & 2.53   & 0.70   \\
 40  & 7 & 0.9464   & 29.90   & 16.21   & 11.02   & 4.50   & 3.06   & 0.85   \\
 45  & 8 & 1.1330   & 35.80   & 19.41   & 13.19   & 5.38   & 3.66   & 1.01   \\
 50  & 9 & 1.3490   & 42.62   & 23.11   & 15.71   & 6.41   & 4.36   & 1.21   \\
 55  &10 & 1.6020   & 50.62   & 27.45   & 18.65   & 7.61   & 5.17   & 1.43   \\
 60  &11 & 1.9030   & 60.13   & 32.60   & 22.16   & 9.04   & 6.14   & 1.70   \\
 65  &12 & 2.2690   & 71.69   & 38.87   & 26.42   & 10.78  & 7.33   & 2.03   \\
 70  &13 & 2.7240   & 86.07   & 46.67   & 31.72   & 12.94  & 8.79   & 2.44   \\
 75  &14 & 3.3060   & 104.45  & 56.64   & 38.49   & 15.70  & 10.67  & 2.96   \\
 80  &15 & 4.0870   & 129.13  & 70.02   & 47.59   & 19.41  & 13.19  & 3.66   \\
 85  &16 & 4.9580   & 156.65  & 84.94   & 57.73   & 23.55  & 16.01  & 4.44   \\
 & & & & & & & & \\
\hline
\hline
\end{tabular*}
\end{center}
\label{tab3}
\normalsize
\end{table*}

Since the observational data for the gradients do not extend beyond  $\sim 2
R_{L,e}$ (see Carollo \& Danziger 1994a,b), our models are limited to the
first eleven regions of the galaxy, i.e. to fractionary radii $R/R_{L,e} =
1.6$ or equivalently  the inner sphere containing 55\% of the total luminous
mass $M_{L,T,12}$. Care must be paid when comparing  integrated observational
quantities, such as magnitudes and colours (see below), with model results.

%%%%%%%%%%%%%%%%%%%%%%%%%%%%%%%%%%%%%%%%%%%
\section{Galactic Winds } 

Baum (1959) first discovered that  elliptical galaxies obey a mean 
color-magnitude relation (CMR): colours get redder at increasing luminosity. 
Long ago Larson (1974) postulated that the present-day CMR could be the result 
of {\it  galactic winds} powered by supernova explosions thus initiating a long 
series of chemo-spectro-photometric models of elliptical galaxies standing on 
this idea (Saito 1979a,b; Matteucci \& Tornamb\'e 1987; Arimoto \& Yoshii 1987;
Angeletti \& Giannone 1990; Mihara \& Tahara 1994; Matteucci 1994; Bressan et 
al. 1994; Tantalo et al. 1996; Gibson 1994, 1996, 1997; Gibson \& Matteucci 
1997, and references therein). In brief, gas is let escape from the galaxy and
star formation is supposed to halt when the total thermal energy of the gas 
equates its gravitational binding energy.
 
The same scheme is adopted here, however with minor modifications due to the 
overall properties of the models and the present formalism.

The thermal energy of the gas is sum of three contributions, namely type I and 
II supernovae and stellar winds from massive stars: 

%%%%%%%%%%%%%Equation 33
\begin{displaymath}
E_{th}(r_{j/2},t) = 
\end{displaymath}

\begin{equation}
E_{th}(r_{j/2},t)_{SNI} + E_{th}(r_{j/2},t)_{SNII} + E_{th}(r_{j/2},t)_{W} 
\label{Eth_tot}
\end{equation}

\noindent
where:

%%%%%%%%%%%%%Equation 34
\begin{displaymath}
E_{th}(r_{j/2},t)_{SNI} = 
\end{displaymath}

\begin{equation}
\int_{0}^{t} \epsilon_{SN}(t-t') 
            R_{SNI}(r_{j/2},t') \Delta M_{L}(r_{j/2},T_{G}) dt' 
\label{Esni}
\end{equation}

%%%%%%%%%%%%%%%Equation 35
\begin{displaymath}
E_{th}(r_{j/2},t)_{SNII} = 
\end{displaymath}

\begin{equation}
\int_{0}^{t} \epsilon_{SN}(t-t') 
            R_{SNII}(r_{j/2},t') \Delta M_{L}(r_{j/2},T_{G}) dt' 
\label{Esnii}
\end{equation}
\noindent
and

%%%%%%%%%%%%%%%Equation 36
\begin{displaymath}
E_{th}(r{j/2},t)_{W} = 
\end{displaymath}

\begin{equation}
\int_{0}^{t} \epsilon_{W}(t-t') 
                R_{W}(r_{j/2},t') \Delta M_{L}(r_{j/2},T_{G}) dt' 
\label{Ew}
\end{equation}

\noindent
with obvious meaning of the symbols. As the production rates 
$R_{SNI}(r_{j/2},t)$, $R_{SNII}(r_{j/2},t)$ and $R_{W}(r_{j/2},t)$ 
are the same as in the set of equations governing the chemical evolution, which
are expressed as a function of the dimensionless variables 
$G_{g,i}(r_{j/2},t)$, the normalization factor $\Delta M_{L}(r_{j/2})$ in the 
equations above is required to calculate the energy in physical units.

The time $t'$ is either the SN explosion time or the time of ejection of the
stellar winds as appropriate. The functions $\epsilon_{SN}(t)$ and
$\epsilon_{W}(t)$ are cooling laws governing the energy content of supernova 
remnants and stellar winds, respectively.

Finally, shell by shell, star formation and chemical enrichment are halted,
and the remaining gas content is supposed to be expelled out of the galaxy 
(winds) when the condition

%%%%%%%%%%%%%Equation 37
\begin{equation}
E_{th}(r_{j/2},t) \geq \Omega_{g}(r_{j/2},t)
\label{eth_omg}
\end{equation}

\noindent
is verified. 

%********************************************************
\subsection{Supernovae and Wind Rates} 

Although the production rates have already been used to define the set of 
equations governing the evolution of the $G_{g,i}(r_{j/2},t)$, they are also 
re-written here for the sake of clarity.

%%%%%%%%%%%%%Equation 38
\begin{displaymath}
R_{SNI}(r_{j/2},t) = ~~~~~~~~~~~~~~~~~~~~~~~~~~~~
\end{displaymath}
\begin{equation}
\Lambda  \int_{M_{Bm}}^{M_{BM}} \phi(M_{B}) 
\int_{\mu_{min}}^{0.5} f(\mu) \Psi(r_{j/2},t-t_{M_2})d\mu dM_{B}
\label{rsni}
\end{equation}

\noindent

%%%%%%%%%%%%%Equation 39
\begin{displaymath}
R_{SNII}(r_{j/2},t) = 
(1-\Lambda) \int_{6}^{16} \phi(M) \Psi(r_{j/2},t-t_{M}) dM ~~~~~
\end{displaymath}

\begin{equation}
~~~~~~~~~~~~~~~~ + \int_{16}^{M_{max}} \phi(M) \Psi(r_{j/2},t-t_{M}) dM 
\label{rsnii}
\end{equation}

\noindent
and finally

%%%%%%%%%%%%%Equation 40
\begin{equation}
R_{W}(r_{j/2},t) = \int_{30}^{M_{max}} \phi(M) \Psi(r_{j/2},t-t_{M}) dM 
\label{rw}
\end{equation}

\noindent
The meaning of all the symbols is the same as above.

%**************************************************
\subsection{Evolution of supernova remnants}

In this section we briefly summarize the prescription we have adopted for the 
cooling law of supernova remnants and the final energy deposit. The formulation
strictly follows Gibson (1994, 1996 and references therein).

The evolution of a SNR can be characterized by three dynamical phases:
(i) free expansion (until the mass of the swept up interstellar material 
reaches that of the SN ejecta); (ii) adiabatic expansion  until the radiative 
cooling time of newly shocked gas equals the expansion time of the remnant; 
(iii) formation of a cold dense shell (behind the front) which begins when 
some sections of the shocked gas have radiate most of their thermal energy, 
begin further compressed by the pressure of the remainder of the shocked 
material.)

In the earliest phase the evolution of the supernova remnant is governed by 
the Sedov-Taylor solution for a self-similar adiabatic shock (Ostriker \& 
McKee 1988)

%%%%%%%%%%%%%Equation 41
\begin{equation}
R_{s}(t) = 1.15 \left(\frac{E_{0}}{\overline{\rho}_{g}(t)}\right)^{1/5} ~~
t^{2/5}
\label{rs}
\end{equation}

\noindent
where $R_{s}(t)$ is the radius of the outer edge of the SNR shock front, 
$E_{0}$ is the initial blast energy in units of $10^{50}~ergs$ (or equivalently
$E_0= 10 \times \epsilon_0$, where $\epsilon_0$ is  the same energy in units of
$10^{51}$ erg), and $\overline{\rho}_{g}(t)$ is the gas mass density of the 
environment.

Radiative cooling of the shocked material leads to the formation of a thin,
dense shell at time $t_{sf}$

%%%%%%%%%%%%%Equation 42
\begin{equation}
t_{sf} = 3.61 \times 10^{4} ~ \epsilon_{0}^{3/14} ~ n_{0}^{-4/7} ~
        \left({Z \over Z_{\odot}}\right)^{-5/14}   ~~~~{\rm yr}
\label{tsf}
\end{equation}
\noindent
where $n_{0}$ is the hydrogen number density, $Z$ is the metallicity of the 
interstellar medium, and \Zsun = 0.016. The blast wave decelerates until the 
radiative energy lost in the shell's material starts to dominate. At this 
point, the shell enters  the so-called ``pressure-driven snowplow'' (PDS) phase
at the time $t_{pds} \approx 0.37t_{sf}$. 

The evolution of the thermal energy in the hot, dilute interior of the 
supernova remnants can be taken equal to

%%%%%%%%%%%%%Equation 43
\begin{equation}
\epsilon_{SN}(t_{SN}) = 0.717 ~ E_{0}~~~~~~~~~~{\rm erg}
\label{esn1}
\end{equation}

\noindent
when $t_{SN} \leq t_{pds}$ i.e. during the adiabatic phase. Note that $t_{SN}
= t-t'$ is the time elapsed since the supernova explosion. During the early 
PDS-phase, when $t_{pds} \leq t_{SN} \leq 1.17~t_{sf}$, the thermal energy
evolution is given by

%%%%%%%%%%%%%Equation 44
\begin{displaymath}
\epsilon_{SN}(t_{SN}) = 0.29 ~
E_{0} ~\left[1-\left(\frac{0.86 ~ t_{SN}}{t_{sf}}\right)^{14/5}\right] + 
\end{displaymath}
\begin{equation}
 ~ 0.43 ~ E_{0} ~ \left[\left(\frac{R_{s}}{R_{sf}}\right)^{10}
+ 1\right]^{-1/5} \left[\left(\frac{t_{SN}}{t_{sf}}\right)^{4} +
1\right]^{-1/9} ~{\rm erg}
\label{esn2}
\end{equation}

\noindent
and the radius changes according to

%%%%%%%%%%%%%Equation 45
\begin{equation}
R_{s}(t_{SN}) = R_{pds} \left(\frac{4 ~t_{SN}}{3 ~t_{pds}} -
\frac{1}{3}\right)^{3/10} ~~~~ {\rm pc}
\label{rs1}
\end{equation}

\noindent
where $R_{pds}$ is the radius at the beginning of the PDS-stage

%%%%%%%%%%%%%Equation 46
\begin{equation}
R_{pds} = 14. ~ \epsilon_{0}^{2/7} ~ n_{0}^{3/7} ~ 
         ({Z \over Z_{\odot}})^{-1/7}  ~~~~ {\rm pc} 
\label{rpds}
\end{equation}

The interior continues to lose energy by pushing the shell through the 
interstellar medium and by radiative cooling. At time $t_{merge}$

%%%%%%%%%%%%%Equation 47
\begin{equation}
t_{merge} = 21.1 ~ t_{sf} ~ \epsilon_{0}^{5/49} ~ n_{0}^{10/49} ~ 
                ({Z \over Z_{\odot}}) ^{15/49}  ~~~~~ {\rm yr}
\label{tmerge}
\end{equation}

\noindent
the remnants merge with the interstellar medium  and lose their 
identity as separate entities.

The thermal energy during the time interval $1.17t_{sf} \leq t_{SN} \leq
t_{merge}$ is given by the second term of Eq.~\ref{esn2}. The evolution after
the $t_{merge}$ time is described again by the second term of Eq.~\ref{esn2},
but the radius $R_{s}$ is given by

%%%%%%%%%%%%%Equation 48
\begin{equation}
R_{s} = R_{merge} = 3.7 ~ R_{pds} ~ \epsilon_{0}^{3/98} ~ n_{0}^{3/49} ~ 
                ({Z \over Z_{\odot}})^{9/98} ~~{\rm pc}
\label{rs2}
\end{equation}
\noindent
Finally, when $t_{SN} \geq t_{cool}$ in which

%%%%%%%%%%%%%%%%Equation 49
\begin{equation}
t_{cool}=203~ t_{sf}~ ({Z \over Z_{\odot}})^{-9/14} ~~~~{\rm  yr}
\label{tcool}
\end{equation}

\noindent
the thermal energy is given by the second term of Eq.~\ref{esn2} but with
$R_{s} = R_{merge}$.

\begin{figure}
%\picplace{9cm}
%\psfig{file=f_cool_sup.ps,height=9.0truecm,width=8.5truecm}
\psfig{file=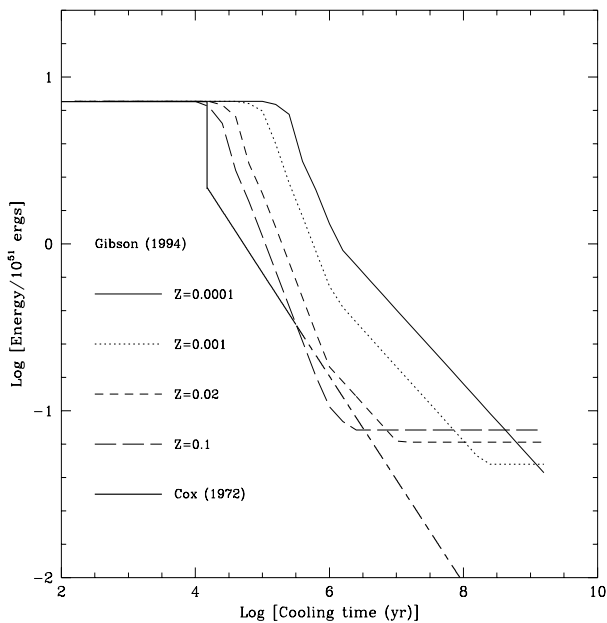,height=9.0truecm,width=8.5truecm}
\caption{The cooling law of supernova remnants as a function of the gas
metallicity as indicated}
\label{f_su_no}
\end{figure}

The time dependence of the cooling law for interstellar media with different
metallicities is shown in Fig.~\ref{f_su_no} and is compared with the
classical one by Cox (1972). It is soon evident that this more elaborated
scheme for the cooling of supernova remnants  supplies more energy to the
interstellar medium than the old one. The adoption of the  Cox (1972) cooling
law by Bressan et al. (1994) and TCBF96 may also explain why they had to
invoke other sources of energy to power galactic winds (see the remark below).

%%%%%%%%%%%%%%Figure 6

%******************************************************
\subsection{Thermal content of the winds}

The thermal content in the winds ejected by massive stars is estimated in the
following way. A typical massive star (say in the range 20\Msun~ to 120\Msun)
in the course of evolution ejects about half of its mass in the form of a wind
with terminal velocity of about 2000 km~$\rm sec^{-1}$ (see Chiosi \& Maeder 
1986) and therefore injects into the interstellar medium an energy of about

%%%%%%%%%%%%%Equation 50
\begin{equation}
\epsilon_{W0} = \eta \frac{1}{2} \left(\frac{M}{2}\right)\left(
\frac{Z}{Z_{*}}\right)^{0.75} V^{2}~~~~~~~{\rm erg}
\label{ewo}
\end{equation}

\noindent
where the term $(Z/Z_{*})^{0.75}$ takes into account that both 
mass loss rates and terminal velocities depend on  metallicity (Kudritzki
et al. 1987, Theis et al. 1992), and $\eta$ is the efficiency parameter 
of kinetic energy thermalization. The reference metallicity is $Z_{*} =0.02$. 

The evolution of adiabatic interstellar bubbles as a result of stellar winds
interacting with the surrounding medium, including radiative losses, confines 
the efficiency parameter in the range $0.2 \leq \eta \leq 0.4$ (Weaver et al. 
1977; Koo \& McKee 1992; Gibson 1994). We have assumed $\eta=0.3$. 

By analogy with the formalism used to calculate the residual thermal energy of 
supernova remnants as a function of time, we write 

%%%%%%%%%%%%%Equation 51
\begin{displaymath}
\epsilon_{W}(t_{W}) =
              \epsilon_{W0}  ~~~~~~~~~~~~~~~~~~~~ 
               { {\rm if}~~~ 0 \leq t_{W} \leq t_{c\omega}} 
\end{displaymath}
or

\begin{equation}
\epsilon_{W}(t_{W}) = \epsilon_{W0} (\frac{t_{W}}{t_{c\omega}})^{-0.62} 
                  ~~~~~ { {\rm if}~~~ t_{W} \geq t_{c\omega} } 
\label{ew}
\end{equation}

\noindent
where $t_{W} = t-t'$ is the time elapsed since the birth of a massive star,
$t_{c\omega}$ is  the cooling time scale. When stellar winds were first 
introduced by Bressan et al. (1994) in the calculation of the total thermal 
budget, the following parameters were adopted: $\eta=1$ and 
$t_{c\omega}=15\times 10^6$ yr. This latter in particular was conceived as the 
typical lifetime of a newly born group of massive stars (either in clusters or 
associations) able to lose mass at a significant rate (indeed $t_{cw}=15\times 
10^{6}$ yr is the typical lifetime of a $10 M_{\odot}$ object).  Bressan's et 
al. (1994) approach  did not pass Gibson' (1994) scrutiny who correctly 
pointed out that only a fraction of the kinetic energy goes thermalized 
($\eta=0.3$) and that $t_{c\omega}$ should be set equal to a star's lifetime
and therefore should  vary with the star mass. Of course the adoption of that
particular set of parameters by Bressan et al. (1994) and later by TCBF96 
led to {\it early } galactic winds as compared to the significantly {\it later} 
winds found by Gibson (1994). In a subsequent paper along the same vein, 
Gibson (1996) suggested that part of the reason why Bressan et al. (1994) 
looked for additional sources of energy (the stellar winds) in addition to 
supernov\ae\ in order to avoid saturation in the metallicity and failure in 
matching the CMR resided in a mismatch of the stellar yields of metals in their
chemical code. Since our goal is not to argue against Gibson's criticism, nor 
to embark in a {\it vis-a-vis} comparison of the codes, in the mean time the 
chemical code has been fully revised and up-graded with respect to the old one,
and the arguments given by Gibson (1994, 1996) are convincing, we definitely 
follow his favourite prescription: $\eta=0.3$ and $t_{c\omega}$ shorter than
mean lifetime of the most massive stars contributing to stellar wind energies. 
In the models below $t_{c\omega}= 10^6$ yr, see Fig.2 in Gibson (1994).

%%%%%%%%%%%%%%%%%%%%%%%%%%%%%%%%%%%%%%%%%%%%%%%%
\section{General properties of the models}

The main properties of each  model at the stage of galactic wind are
summarized in Table 4 (not given here but available from the A\&A electronic
data-base) as a function of the radial distance from the center. Column (1) is
the asymptotic mass $M_{L,T,12}$, column (2) is the efficiency of star
formation $\nu(r_{j/2})$; column (3) is the IMF parameter $\zeta$; column (4)
is  the time scale of mass accretion $\tau(r_{j/2}$) in Gyr. Column (5) shows
the value reached by $M_{L,t}$ at the onset of galactic wind. This is the real
luminous mass of the galaxy built up by the  infall process, all the remaining
gas (both already accreted and still on the way) being swept away by galactic
winds. Columns (6) through (8) are the age in Gyr at which the galactic wind
occurs, and the corresponding dimensionless mass of gas $G(r,t)$ and living
stars $S(r,t)$, respectively. According to their definition, in order to
obtain the real mass in gas and stars (in solar units) one has to multiply
them by the normalization mass of each shell, i.e. $\Delta M_L(r_{j/2}, T_G)$.
Likewise, to get from $G(r,t)$ and $S(r,t)$ the corresponding densities, the
multiplicative factor is $\overline{\rho}_L(r_{j/2}, T_G)$. Columns (9) and
(10) are the maximum and mean metallicity, $Z(r_{j/2},t)$ and $\rm \langle
Z(r_{j/2},t) \rangle$ reached by each shell at the onset of the  galactic
wind. Column (11) contains the rate of star formation $\Psi(r_{j/2},t)$ in
units of $M_{\odot}$/yr. Columns (12) through (16) are  the gravitational
binding energy of the gas $\Omega_{g}(r_{j/2},t)$, the total thermal energy of
this $E_{g}(r_{j/2},t)$, and the  separated contributions by Type I, Type II
supernovae, and stellar winds, respectively. All energy are in units of
$10^{50}~\rm ergs$. Finally, column (17) is the mid shell fractionary radius
$r_{j/2}$.

%***************************************************
\subsection{Internal consistency of the models}
The scheme we have elaborated in the previous sections is self-contained in 
absence of galactic winds, because in such case at the galaxy age $T_G$ all
shells have reached their asymptotic mass and the effective radius 
$R_{L,e}$ (the basic scale factor associating the asymptotic density of the 
Young profile to each radius) is consistent with $M_{L,T}(T_G)$. 

At the stage of galactic wind we suppose that all the gas contained in the 
shell, the one still in the infall process and the one expelled by supernova 
explosions and stellar winds are ejected into the intergalactic medium and 
never re-used to form stars. This implies that at the stage of galactic wind 
the real mass of each shell (the fraction of gas turned into long-lived stars 
up to this stage), is smaller than the corresponding asymptotic mass 
$\Delta M_{L}(r_{j/2},T_G)$. Indeed in each shell the luminous mass has grown 
up to the value  $\Delta M_{L}(r_{j/2},t_{gw})$, where $t_{gw}$ is the local 
value of the age at the onset of the galactic wind. Therefore

%%%%%%%%%% equation 52
\begin{equation}
            \Sigma_{j=0}^{J-1} \Delta M_{L}(r_{j/2},t_{gw})  < M_{L,T}(T_G)
\label{sum_mass_parz}
\end{equation}

Recalling that our calculations refer to the innermost part of the galaxy (the 
one containing 55\% of the mass $M_{L,T}(T_G)$), the relation 
(\ref{sum_mass_parz}) should be replaced by 

%%%%%%%%%% equation 53
\begin{equation}
            \Sigma_{j=0}^{10} \Delta M_{L}(r_{j/2},t_{gw})  < 0.55
                           \times M_{L,T}(T_G)
\label{sum_mass_parz_1}
\end{equation}

%%%%%%%%Table 4 Not given here %%%%%%%%%%%%%%%

%%%%%%%%%Table 5  masses and radii %%%%%
\setcounter{table}{4}
\begin{table*}
\vskip 0.3 cm
\begin{center}
\vskip 0.1 cm
\caption{Fractionary masses of gas and stars components in units of 
$10^{12}$\Msun for the models presented in this work.}
\vskip 0.1 cm
\scriptsize
\begin{tabular*}{130mm} {r r c c c r }
\hline
\hline
 & & & & & \\
 $M_{L,T}(T_G)$ &  $R_{L,e}(T_G)$ &  $M_{L}(1.5 R_{L,e},T_G)$ & 
 $M_{L,T}(1.5 R_{L,e}, t_{gw})$ &  $M_{L}(R_{L,e},t_{gw})$  
  & $R_{L,e}(t_{gw})$   \\
 & & & & & \\
\hline
 & & & & & \\
 3     & 31.6 & 1.65000 & 0.65700 & 0.65600 & 13.55  \\
 1     & 17.1 & 0.55000 & 0.21800 & 0.21500 &  7.33  \\
 0.5   & 11.7 & 0.27500 & 0.10900 & 0.10700 &  4.98  \\
 0.1   & 4.7  & 0.05500 & 0.02200 & 0.02100 &  2.04  \\
 0.05  & 3.2  & 0.02750 & 0.01090 & 0.01070 &  1.38  \\
 0.005 & 0.9  & 0.00275 & 0.00102 & 0.00097 &  0.37  \\
 & & & & & \\
\hline
\hline
\end{tabular*}
\end{center}
\normalsize
\label{tab5}
\end{table*}

Looking at the case of the 3$M_{L,T,12}$ galaxy, the sphere we have been
following in detail has total asymptotic  mass of $1.65 \times 10^{12}
M_{\odot}$, each shell containing about $0.15\times 10^{12} M_{\odot}$ (cf.
Column (5) of Table 4). In contrast, the total mass reached in the same sphere
at the onset of the galactic wind amounts only to $0.66 \times 10^{12} 
M_{\odot}$, i.e. some 40\% of the expected mass. Even more important, while the
innermost shells were able to convert in stars about 0.8 of their asymptotic 
mass, this is not the case of the outermost shells in which only about 2\% of 
the potential mass has been turned into stars, all the rest being dispersed by 
a very early wind. Considering that owing to the very low densities in regions
above our last shell (approximately $1.5 R_{L,e}$) the galactic winds
would occur even earlier than in the last computed shell, this means that
starting with 3$M_{L,T,12}$ of gas eligible to star formation only 22\% of
it has been actually turned into long-lived stars visible today. The situation
gets slightly better at decreasing $M_{L,T}(T_G)$ because of the much shorter
mean infall time scale (cf. Tables 5 and 1).

Furthermore, if we look at the radial profile of $\overline{\rho}_L(r_{j/2},
t_{gw})$ and compare it with $\overline{\rho}_L(r_{j/2}, T_G)$, the former is
steeper than the latter,  over the shells external $R_{L,e}$ in particular.
However, if we limit the comparison to the shells inside $R_{L,e}$ (up to
$j=8$ in our notation), the difference is remarkably smaller. This implies
that the region inside $R_{L,e}$ does not depart too much from the basic
hypothesis. Finally, the effective radius $R_{L,e}$ used to interpolate in the
Young (1976) density profile and to assign $\overline{\rho}_L(r_{j/2},T_G)$
referred to the asymptotic mass $M_{L,T}(T_G)$. Since the actual present-day
mass of the galaxy is smaller than this, we expect the actual effective radius
to be smaller than the originally adopted value. With the aid of relation
(\ref{reff_mass}) above, the 3$M_{L,T,12}$ galaxy has $R_{L,e}\simeq 31.9$
Kpc, whereas the 0.66$M_{L,T,12}$ daughter should have $R_{L,e}\simeq 13.7$
kpc (a factor 2.3 smaller). This means that the ratio of the mean density
(inside $R_{L,e}$) of the parent to daughter galaxy is about 0.5. It is as if
we calculated our models under-estimating their real density by a factor of
about two. Considering that even within the effective radius passing from the
center to the periphery the density of luminous mass drops by orders of
magnitude, cf. Young (1976), and all other uncertainties affecting our models,
we can perhaps tolerate  the above discrepancy. The results we are going to
present  perhaps constitute the best justification of these models, which do
not dare to replace more sophisticated, physically grounded formulations in
literature, but simply aim at providing  a  simple tool to investigate the
chemo-spectro-photometric properties and their spatial gradients of spherical
systems roughly simulating elliptical galaxies.

Table 5 summarizes the data relative to the above discussion for all the model
galaxies under examination. It lists the asymptotic total mass $M_{L,T}(T_G)$
(column 1), the corresponding effective radius $R_{L,e}(T_G)$ (column 2), the
asymptotic mass $M_{L}(1.5 R_{L,e},T_G)$ within $1.5 R_{L,e}$ (the
studied model, column 3), the actual mass $M_{L,T}(t_{gw})$ of the galaxy
within $1.5 R_{L,e}$ at the age $t_{gw}$ (column 4), the actual mass
$M_{L,T}(R_{L,e},t_{gw})$ of the galaxy within $ R_{L,e}$  at the age $t_{gw}$
(column 5), and the real effective radius $R_{L,e}(t_{gw})$ (column 6).

%*************************************************************
\subsection{Gas content, metallicity, SFR, and N(Z)}

\begin{figure}
%\picplace{9cm}
%\psfig{file=f_ga_ze.ps,height=9.0truecm,width=8.5truecm}
\psfig{file=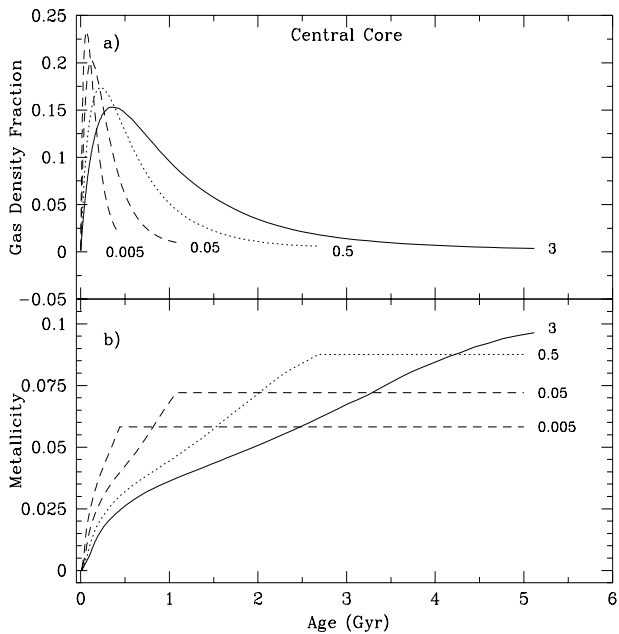,height=9.0truecm,width=8.5truecm}
\caption{The gas fraction $G(r,t)$ ({\em top panel }) and metallicity $Z(r,t)$
({\em bottom panel }) as a function of the age in $Gyr$ for the central core
of the galaxy models with different asymptotic mass $M_{L,T,12}$ as indicated}
\label{f_ga_ze}
\end{figure}

The fractionary gas content $G(r,t)$ and metallicity $Z(r,t)$ for  the central
core of the models (up the fractionary radius $r_{1/2}$) are shown as function
of time  in panels (a) and (b), respectively, of Fig.~\ref{f_ga_ze}. In all
the models, the fractionary gas density  $G_g(r_{1/2},t)$ starts small,
increases up to a maximum and then decreases exponentially to zero as a result
of the combined effect of gas accretion by infall and gas consumption by star
formation, but owing to the  different value of $\tau$ in the core from model
to model, the peak occurs later at increasing galaxy mass $M_{L,T}(T_G)$.

As far as the metallicity is concerned, this increases more slowly at
increasing galaxy mass up to the maximum value reached in coincidence of the
galactic wind. As expected the maximum metallicity increases with the galaxy
mass, because in this type of model galactic winds occur later at increasing
galaxy mass (cf. the entries of Table 4 and Fig.~\ref{f_w_age} below).

%%%%%%%%%%%%%%Figure 7

\begin{figure}
%\picplace{9cm}
%\psfig{file=f_z_zmx.ps,height=9.0truecm,width=8.5truecm}
\psfig{file=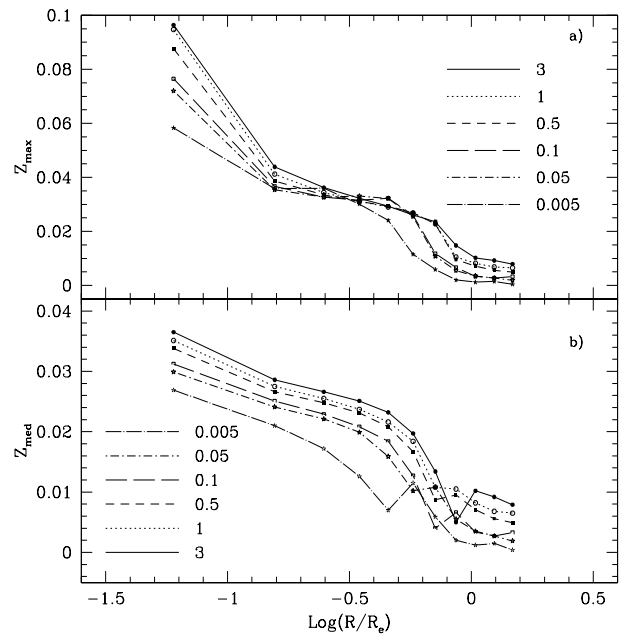,height=9.0truecm,width=8.5truecm}
\caption{The gradients in maximum (top panel) and mean metallicity (bottom 
panel) for the model galaxies with different $M_{L,T,12}$ as indicated}
\label{f_z_zmx}
\end{figure}

Fig.~\ref{f_z_zmx}  shows the  maximum ($Z_{max}$, top panel) and mean
($Z_{mean}$, bottom panel)  metallicity as a function of the radial distance
from the center (normalized to the effective radius of each galaxy) for all
the  models  as in Figs.~\ref{f_tau_ra} and ~\ref{f_nu_ra}. The mean gradient
in the maximum metallicity, $dZ_{max}/dlog(r)$, within $1.5 R_{L,e}$ ranges
from --0.064 to --0.042 going from massive to dwarf galaxies, whereas  the mean
gradient in  mean metallicity, $dZ_{mean}/dlog(r)$, over the same radial
distance and galaxy mass interval goes from --0.021 to --0.019.

%%%%%%%%%%%%%Figure 8

\begin{figure}
%\picplace{9cm}
%\psfig{file=f_sfr_en.ps,height=9.0truecm,width=8.5truecm}
\psfig{file=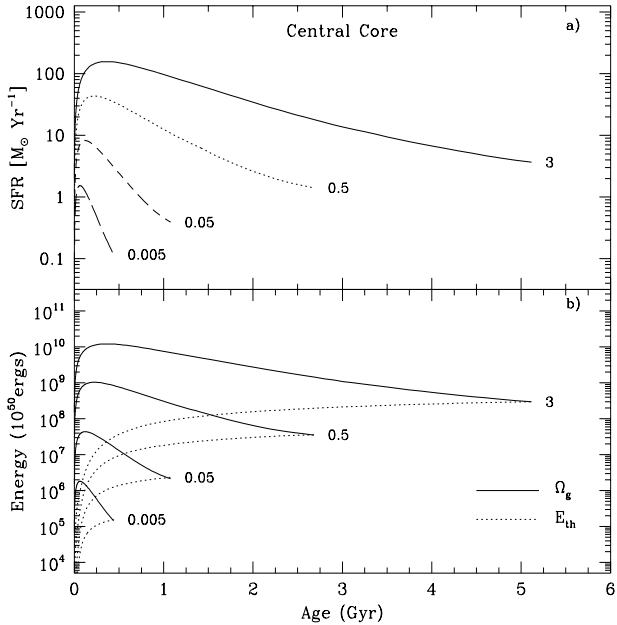,height=9.0truecm,width=8.5truecm}
\caption{{\em Panel a}: the star formation rate as a function of time for the 
central core of the galaxy models with different asymptotic mass $M_{L,T,12}$ 
as indicated. {\em Panel b:} the gravitational binding energy $\Omega_{g}(r,t)$ 
and thermal content of the gas $E_{th}(r,t)$, for the same models as above. 
Energies are in units of $10^{50}~ergs$ }
\label{f_sfr_en}
\end{figure}

The top panel of Fig.~\ref{f_sfr_en} shows the rate of star formation (in units
of \Msun/yr) as a function of time (in Gyr) for the central core of the models,
up to the onset of galactic winds. As expected, the rate of star formation 
starts very small, grows to a maximum, and then declines exponentially with 
time, closely mimicking the gas content. The gas liberated by evolving stars 
(supernova explosions, stellar winds, and PN) in subsequent epochs is not 
shown here as all this gas is supposed to be rapidly heated up to the escape 
velocity.

%%%%%%%%%%%%%%%%Figure 9

The bottom panel of Fig.~\ref{f_sfr_en} displays the comparison between the
thermal and the binding energy of the gas, $E_{th}(r,t)$ and $\Omega_{g}(r,t)$,
respectively, as a function of time for the nuclear regions. All the energies 
are in units of $10^{50}$ erg. The intersection between $\Omega_{g}(r,t)$ and 
$E_{th}(r,t)$ corresponds to the onset of the galactic wind for the innermost 
region. Similar diagrams can be drawn for all the remaining shells. They are 
not displayed for the sake of brevity.

In this type of model galactic winds occur earlier passing from the center 
to external regions, or at given relative distance from the center,  going 
from massive to low mass galaxies. This is shown  in Fig.~\ref{f_w_age}  
which displays  the age of the galactic wind  $t_{gw}$ as a function of the 
galacto-centric distance. The stratification in metallicity, and relative 
percentage of stars in different metallicity bins resulting from the above 
trend in $t_{gw}$ bears very much on inferring chemical abundances from local 
or integrated photometric properties of elliptical galaxies. This topic will 
be addressed below in some detail.

%%%%%%%%%%%%%%%%Figure 10
\begin{figure}
%\picplace{9cm}
%\psfig{file=f_w_age.ps,height=9.0truecm,width=8.5truecm}
\psfig{file=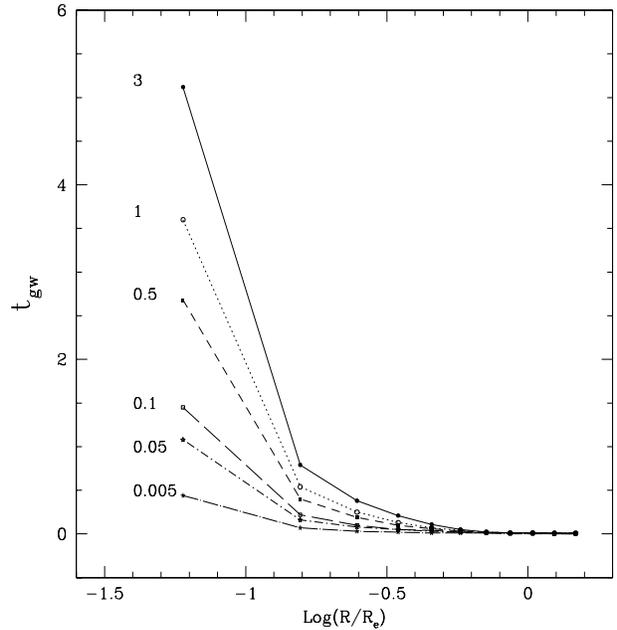,height=9.0truecm,width=8.5truecm}
\caption{ The age at which galactic winds occur in regions of increasing 
distance from the galactic centre. The models are the same as in 
Figs.~\ref{f_tau_ra} and \ref{f_nu_ra}.}
\label{f_w_age}
\end{figure}

The chemical structure of the models is best understood looking at the
fractionary cumulative mass distribution of living stars, $\Sigma _0^Z S_Z/S$,
where $S$ is the mass fraction in stars, and $S_Z$ is the mass fraction of
stars with metallicity up to $Z$, and at the so-called {\it partition
function} $N(Z)$, i.e. the relative number of living stars per metallicity
bin. Within a galaxy (or region of it) both distributions vary as a  function
of the age.

\begin{figure}
%\picplace{9cm}
%\psfig{file=f_ms_ze.ps,height=9.0truecm,width=8.5truecm}
\psfig{file=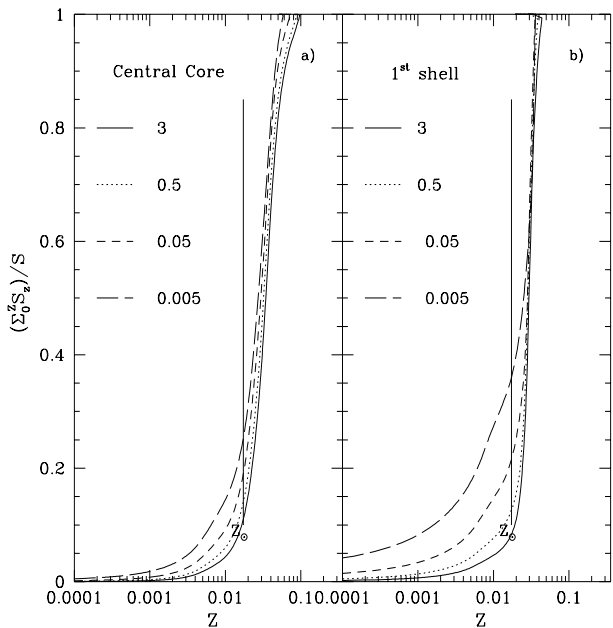,height=9.0truecm,width=8.5truecm}
\caption{The cumulative fractionary mass of living stars as a function of the
metallicity for the galaxy models with mass 3, 0.5, 0.05, and 0.005 
$M_{L,T,12}$. {\em Panels a} and $b$ corresponds to the central core and first 
shell respectively}
\label{f_ms_ze}
\end{figure}

The fractionary, cumulative mass distribution as a function of $Z$  is shown
in Fig.~\ref{f_ms_ze} limited to the central core ($r_{1/2}$, left panel) and
the first shell ($r_{3/2}$, right panel) for all the models  at the present 
age ($T_G=15$ Gyr). The vertical line corresponds to the solar metallicity. In
the core and the first shell of the most massive galaxy, about 10\% of the 
stars have metallicity lower than solar. In contrast, the central region of the
lowest mass galaxy has about 25\% of its stellar content with metallicity lower
than solar. This percentage increases to about 36\% in the first shell. In all 
galaxies the percentage of stars with metallicity lower than solar increases 
as we move further out.

%%%%%%%%%%%%%%%%Figure 11

\begin{figure}
%\picplace{9cm}
%\psfig{file=f_ns_zb.ps,height=9.0truecm,width=8.5truecm}
\psfig{file=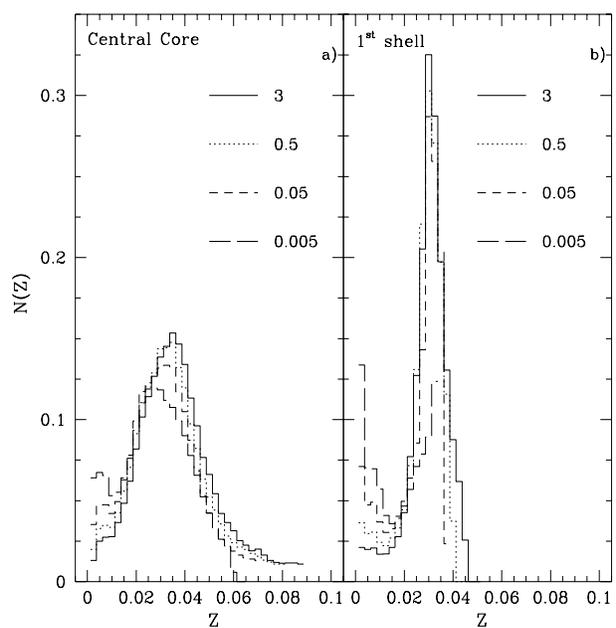,height=9.0truecm,width=8.5truecm}
\caption{Relative number of living stars per metallicity bin in the central
core {\em panel a)} and first shell {\em panel b)} for the models with 3, 0.5 0.05 
and 0.005 $M_{L,T,12}$ (the same models as in Fig.~\ref{f_ms_ze})}
\label{f_ns_zb}
\end{figure}

The partition function $N(Z)$ for our model galaxies at the age of 15 Gyr is 
shown in Fig.~\ref{f_ns_zb} limited to the central core (left panel) and first 
shell (right panel). We learn from this diagram that the mean metallicity of 
the stars in the core goes from $Z\simeq 0.03$ to $Z\simeq 0.04$; the peak 
value tends to slightly shift toward higher metallicities at increasing galaxy 
mass; and there are wings toward both low and high metallicities. The 
distribution tends to be more concentrated in the first shell, where we notice 
a more abundant population of low metallicity stars and a sharper cut-off at 
the high metallicity edge caused by the action of galactic winds. Likewise for 
the remaining shells not displayed here. 

%%%%%%%%%%%%%%%%Figure 12

%%%%%%%%%%%%%%%%%%%%%%%%%%%%%%%%%%%%%%%%%%%%%%%%%%
\section{Photometric Properties}

As already mentioned, the guide-line for the layout of the model and the
choice of the various parameters was to impose that a number of properties of 
elliptical galaxies could be simultaneously matched. Specifically: (i) the 
slope of color-magnitude relation (CMR) (Bower et al. 1992a,b); (ii) the mean 
value of the broad-band colours; (iii) the UV excess as measured by the colour
(1550--V) (Burstein et al. 1988); (iv) the mass to blue luminosity ratio
$(M/L_{B})_{\odot}$ as a function of the B luminosity (Bender et al. 1992, 
1993); (v) the $R^{1/4}$ luminosity profile (Fasano et al. 1996); (vi) the 
gradient in (B-R) colour measured by  Carollo \& Danziger (1994a,b) in a 
sample of elliptical galaxies; (vii) the gradients in line strength indices 
$Mg_2$ and $\rm \langle Fe \rangle$ measured by Carollo \& Danziger (1994a,b). 
(viii) and finally, the data of Gonz\'ales (1993) for the ${H_{\beta}}$ and 
$[MgFe]$ line strength indices. In this section, we present the comparison of 
model results with the observational data in relation to the above list of 
observational {\it constraints}.

To this aim we need to calculate the integrated colours and line strength
indices together with their gradients for the stellar mix in the 
model galaxies. The technique in usage here is based on the concept of single
stellar populations (SSP) as  elemental seeds to derive the integrated stellar
energy distribution (ISED) of a galaxy, from which magnitudes, broad-band
colours, and line strength indices immediately follow. The SSP's adopted in
this paper are those calculated by TCBF96, see also Bressan et al. (1994, 
1996), to whom the reader should refer for technical details. First we have 
calculated the integrated  ISED, magnitudes, colours etc. for each zone of our 
models, and then we have derived the total magnitudes, colours etc. for the 
whole galaxy.

Table 6 (not given here but available in A\&A electronic data-base) lists the
integrated magnitudes and broad-band colours of every zone of the model
galaxies at three different ages (15, 10, and 5 Gyr). Columns (1) through (5)
display: the asymptotic mass of the model (in units of $10^{12}$\Msun),
$\nu(r_{j/2})$, $\tau(r_{j/2})$ (in Gyr), the asymptotic mass of each zone
(in units of $10^{12}$\Msun) and the age in Gyr, respectively. Columns (6) and
(7) give the integrated absolute bolometric ($M_{bol}$) and visual magnitudes
($M_{V}$) of each zone, respectively. Columns (8) through (15) are the
integrated colours (U--B), (B--V), (V--R), (V--I), (V--J), (V--H), (V--K), and
(1550--V). Finally, Column (16) gives the fractionary radii $r_{j/2}$.

Table 7 shows the same quantities but integrated from the center up the $1.5
R_{L,e}$ radius. These are the quantities to be used to compare theory  with
observations.

%%%%%%%%%%Table 6 not given here 

%%%%%%%%%%%%%Table 7 Integrated magnitudes and colours
\setcounter{table}{6}
\begin{table*}
\begin{center}
\caption{Integrated magnitude and colours form the center up the $1.5 
R_{L,e}$-sphere of the model galaxies}
\scriptsize
\begin{tabular*}{155mm}{r r c  c c c  c c c  c c c  c  }
\hline
\hline	 
 & & &  & & &  & & &  & & &  \\
			 \multicolumn{1}{r}{$M_{L,T,12}$}  
			&\multicolumn{1}{r}{Age}
			&\multicolumn{1}{c}{$M_{bol}$}
			&\multicolumn{1}{c}{$M_{V}$}
			&\multicolumn{1}{c}{(U--B)}
			&\multicolumn{1}{c}{(B--V)}
			&\multicolumn{1}{c}{(V--R)}
			&\multicolumn{1}{c}{(V--I)}
			&\multicolumn{1}{c}{(V--J)}
			&\multicolumn{1}{c}{(V--H)}
			&\multicolumn{1}{c}{(V--K)}
			&\multicolumn{1}{c}{(1550--V)}
			&\multicolumn{1}{c}{$r_{j/2}$} \\ 
 & & &  & & &  & & &  & & &  \\ 
\hline   
(1)&(2)&(3)&(4)&(5)&(6)&(7)&(8)&(9)&(10)&(11)&(12)&(13)\\
\hline
  & & &  & & &  & & &  & & &  \\
3.0& 15 &-23.536 &-22.700 &0.557 &0.991 &0.596 &1.201 &2.396 &3.117 &3.323 &3.135 &1.48\\
3.0& 10 &-23.925 &-23.071 &0.502 &0.962 &0.582 &1.180 &2.432 &3.161 &3.379 &4.762 &1.48\\
3.0&  5.&-24.697 &-23.798 &0.300 &0.845 &0.538 &1.112 &2.440 &3.176 &3.412 &1.283 &1.48\\
  & & &  & & &  & & &  & & &  \\
1.0& 15 &-22.316 &-21.512 &0.515 &0.972 &0.588 &1.186 &2.348 &3.064 &3.264 &3.279 &1.48\\
1.0& 10 &-22.697 &-21.878 &0.461 &0.946 &0.575 &1.166 &2.379 &3.103 &3.314 &4.237 &1.48\\
1.0&  5.&-23.357 &-22.536 &0.368 &0.882 &0.545 &1.118 &2.384 &3.115 &3.338 &5.652 &1.48\\ 
  & & &  & & &  & & &  & & &  \\
0.5& 15 &-21.563 &-20.786 &0.484 &0.957 &0.581 &1.174 &2.309 &3.021 &3.217 &3.404 &1.48\\
0.5& 10 &-21.940 &-21.147 &0.434 &0.936 &0.570 &1.155 &2.340 &3.060 &3.267 &4.257 &1.48\\
0.5&  5 &-22.581 &-21.790 &0.348 &0.873 &0.540 &1.107 &2.340 &3.067 &3.284 &5.506 &1.48\\ 
 & & &  & & &  & & &  & & &   \\
0.1& 15 &-19.802 &-19.101 &0.389 &0.906 &0.560 &1.134 &2.192 &2.888 &3.070 &3.490 &1.48\\
0.1& 10 &-20.170 &-19.455 &0.350 &0.898 &0.552 &1.121 &2.223 &2.928 &3.121 &4.561 &1.48\\
0.1&  5 &-20.784 &-20.076 &0.281 &0.834 &0.523 &1.072 &2.212 &2.923 &3.125 &5.078 &1.48\\ 
 & & &  & & &  & & &  & & &  \\
0.05& 15&-19.059 &-18.392 &0.350 &0.885 &0.551 &1.116 &2.141 &2.828 &3.006 &3.512 &1.48\\
0.05& 10&-19.425 &-18.742 &0.315 &0.881 &0.544 &1.106 &2.173 &2.869 &3.057 &4.769 &1.48\\
0.05&  5&-20.034 &-19.358 &0.254 &0.818 &0.515 &1.058 &2.160 &2.862 &3.058 &4.916 &1.48\\ 
 & & &  & & &  & & &  & & &  \\
0.005&15 &-16.404&-15.813 &0.266 &0.829 &0.527 &1.070 &2.016 &2.679 &2.843 &3.456 &1.48\\
0.005&10 &-16.762&-16.159 &0.236 &0.837 &0.524 &1.066 &2.047 &2.719 &2.891 &5.151 &1.48\\
0.005& 5 &-17.359&-16.766 &0.189 &0.774 &0.495 &1.017 &2.023 &2.701 &2.881 &4.544 &1.48\\
 & & &  & & &  & & &  & & &  \\
\hline
\end{tabular*}
\end{center}
\normalsize
\label{tab7}
\end{table*}

%***********************************************
\subsection{Color-Magnitude relation}

The CMR for the models in Tables 6 and 7 is compared with the data by Bower et
al. (1992a,b) for the Virgo and Coma elliptical galaxies. Since the
observational data refer to the whole galaxies, the theoretical results of
Table 7 must by suitably corrected to take into account the contribution from
all the other regions not considered here. To this aim we proceed as follows:
the  integrated magnitudes (Table 7) refer to the sphere of $1.5\times R_{L,e}$
in which 55\% of the total mass is contained; the models supposedly obey the 
$R^{1/4}$ law (see also the discussion below); with the aid of items (i) and 
(ii) we calculate the fraction of light coming from the regions from 
$1.5\times R_{L,e}$ to $\infty$. To a first approximation the total magnitudes 
are simply given by

%%%%%%%%% equation 54
\begin{equation}
     M_{\Delta \lambda, T} = M_{\Delta \lambda, (1.5\times Re) } - 0.3342
\end{equation}

\noindent
where $\Delta\lambda$ indicates the pass-band and all the other symbols are
self-explanatory.

\begin{figure}
%\picplace{9cm}
%\psfig{file=f_cmr.ps,height=9.0truecm,width=8.5truecm}
\psfig{file=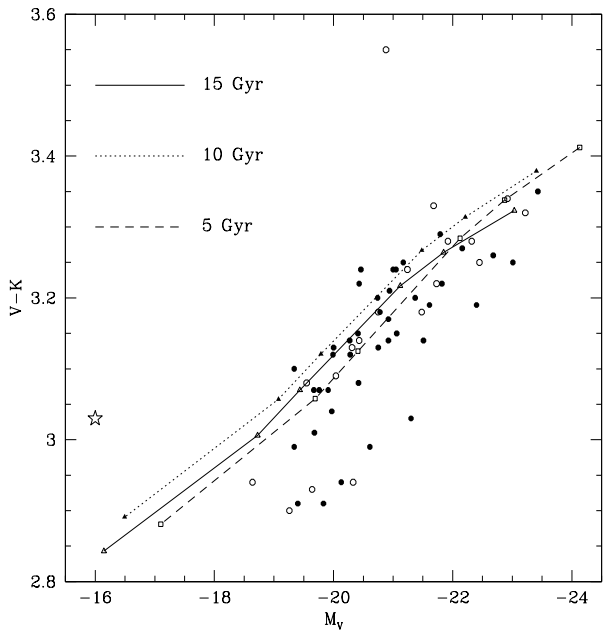,height=9.0truecm,width=8.5truecm}
\caption{The CMR for models with mass of 3, 1, 0.5, 0.1, 0.05, and 0.005 
$M_{L,12}$ and different ages as indicated. The data for the galaxies in Virgo 
({\em open circles}) and Come ({\em filled circles}) are by Bower et al.\ (1992a,b). 
The {\em star} is the galaxy M32.}
\label{f_cmr}
\end{figure}

The comparison between theory and observations is shown in  Fig.~\ref{f_cmr}.
for three values of the age as indicated. The Virgo and Coma galaxies are
displayed with different symbols: open and filled circles, respectively. The
absolute magnitudes V are calculated assuming  the distance modulus to Virgo of
$(m-M)_{o} = 31.54$ (Branch \& Tammann 1992) and applying to the Coma galaxies
the shift $\delta (m-M)_{o} = 3.58$ (Bower et al. 1992a,b).

The agreement is remarkably good both as far as the absolute colours and the
slope of the CMR are concerned. According to Bower et al. (1992a,b) the
thickness of the Virgo-Coma CMR in the (U-V) versus $M_V$ plane implies that
elliptical galaxies in these clusters are old with little age dispersion say
$13\div 15$ Gyr. Although this conclusion is compatible with the data in
Fig.~\ref{f_cmr}, our isochrones in the (V-K) versus $M_V$ plane span a small
range in colour passing from 5 to 15 Gyr, so that confirmation of an old age
from this side is not possible.

%%%%%%%%%%%%Figure 13

%***********************************************
\subsection{Mass to blue luminosity ratio}

To calculate the $(M/L_{B})_{\odot}$ ratios for the model galaxies the 
following procedure is adopted: first using the data of Tables 4 and 5 we 
derive the total present-day value of the galactic mass {\it up to the last 
calculated shell}, second we utilize the integrated magnitudes of Table 7 
{\it to evaluate the blue luminosity}. The results are presented in Table 8 which 
lists the age in Gyr (Column 1), the asymptotic mass $M_{L,T,12}$ (Column 2), 
the current mass of the galaxy in units of $10^{12}$\Msun\ (Column 3), the 
blue magnitude $M_{B}$ (Column 4), the blue luminosity $L_{B}$ in solar units  
(Column 5), and finally, the mass to blue luminosity ratio $(M/L_{B})_{\odot}$ 
(Column 6). As long known, on the theoretical side the mass to blue luminosity 
ratio $(M/L_{B})_{\odot}$ is very sensitive to the IMF, i.e. for a 
Salpeter-like case to the slope and $\zeta$ (the fraction of IMF mass stored 
above say $1M_{\odot}$). In this study we have adopted the Salpeter law and
$\zeta=0.5$. With  this assumption, at any given age the models predict
$(M/L_{B})_{\odot}$ ratios that are nearly constant at increasing luminosity 
(mass) of the galaxy, and at fixed galactic mass they increase by a factor of 
about 3 as the age goes from 5 to 15 Gyr.

On the observational side, the $M/L_{B}$ ratios (in solar units) by Bender et 
al. (1992, 1993) and Terlevich \& Boyle (1993) -- scaled to the Hubble 
constant $H_{0} = 50~ {\rm km~sec^{-1}~Mpc^{-1} }$ -- range from 1 to 18. 

\begin{figure}
%\picplace{9cm}
%\psfig{file=f_mlb_rt_new.ps,height=9.0truecm,width=8.5truecm}
\psfig{file=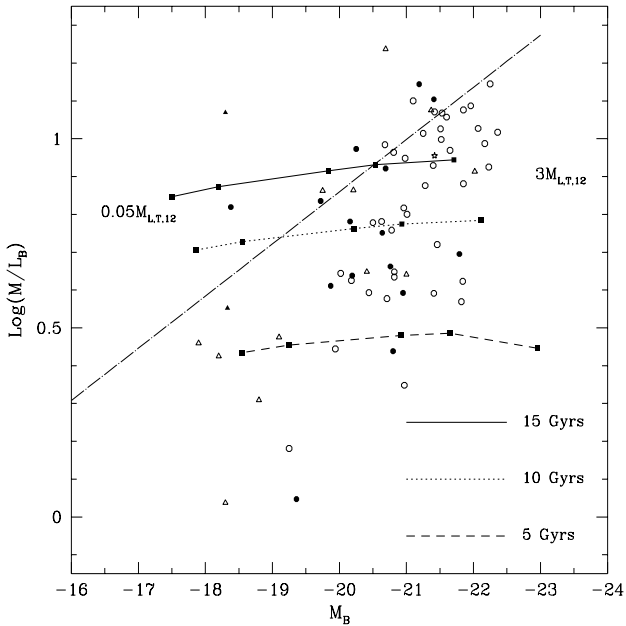,height=9.0truecm,width=8.5truecm}
\caption{The logarithm of the mass to B-luminosity ratio $M/L_{B}$ versus the 
absolute blue magnitude $M_{B}$ for the model galaxies at three different ages,
i.e. 15, 10 and 5 Gyr. The mass used to calculate $M/L_{B}$ and $M_{B}$ refers 
to the present-day mass in form of living stars. The ratio $M/L_{B}$ is 
expressed in solar units. The {\em dashed line} is the relation by Terlevich \& 
Boyle (1993) for $H_{0} = 50 ~{\rm km~sec^{-1}~Mpc^{-1}}$. T
he data are from Bender et
al.\ (1992,1993), i.e. {\em open dots}: giant elliptical's; {\em full dots}: 
intermediate elliptical's; {\em stars}: bright dwarf elliptical's; 
{\em open squares}: compact elliptical's; {\em open triangles}: bulges}
\label{f_mlb_rt}
\end{figure}

The comparison with the observational data is presented in Fig.~\ref{f_mlb_rt} 
for different values of the age as indicated.

Therefore, while the mean values of the mass to blue luminosity ratios agree 
with the data, this type of model is still unable to explain the systematic 
increase of the mass to blue luminosity ratio with the galaxy luminosity for 
coeval, old objects as suggested by the CMR. Possible ways out are: (i) either 
faint galaxies are younger than the bright ones in contrast with the CMR hint 
or (ii) other causes must exist. In relation to this, Chiosi et al. (1998) 
have investigated the possibility that the IMF (cut-off mass and slope) vary 
from galaxy to galaxy in a systematic fashion: the IMF is more top-heavy 
(higher cut-off mass and shallower slope) in the massive elliptical galaxies 
than in the low mass ones (lower cut-off mass and steeper slope). Indeed Chiosi
et al. (1998) models explain the inclination of  the mass to light ratio versus
luminosity (otherwise known as the inclination of the Fundamental Plane).

Before concluding this section we have to check the radial dependence of the 
$(M/L_{B})_{\odot}$ predicted by the models. To this aim we calculate the 
cumulative $(M/L_{B})_{\odot}(r_{j/2})$ moving from the center up to the last 
computed zone. The results are presented in Table~9 limited to a few selected 
radii and the 3 and 0.1 $M_{L,T,12}$ galaxies. The selected radii $r_{j/2}$ 
correspond to the central core, $0.6 R_{L,e}$, $R_{L,e}$ and $2 R_{L,e}$. It is
soon evident that the $(M/L_{B})(r_{j/2})$ ratio is nearly constant (within 
about 10\%) passing from the center to  the external regions. 
{\it This implies that the first condition imposed by the choice of the Young (1976) density profile
for the luminous material, i.e. radially constant mass to luminosity ratio, is almost fully verified. }

%%%%%%%%%%%%%Table 8 {mass to blue-luminosity ratio}
\setcounter{table}{7}
\begin{table}
\vskip 0.3 cm
\begin{center}
\vskip 0.2 cm
\caption{The mass to blue-luminosity ratio (in solar units) as function of the 
age for the model galaxies with different mass.}
\vskip 0.25 cm
\scriptsize
\begin{tabular} {r c c c r c}
\hline
\hline
    & & & & &  \\
Age & $M_{L,T,12}$ & $M_{L,T}(t_{gw})$ & $M_{B}$ & $L_{B}$ 
       & $(M/L_{B})_{\odot}$ \\
    & & & & & \\
\hline
    & & & & & \\ 
 15 & 3     & 0.660 &  -21.71 & 7.509e10 & 8.793 \\
 10 & 3     &       &  -22.11 & 1.085e11 & 6.083 \\
  5 & 3     &       &  -22.95 & 2.362e11 & 2.796 \\
    & & & & & \\
 15 & 1     & 0.218 &  -20.54 & 2.559e10 & 8.540 \\
 10 & 1     &       &  -20.93 & 3.671e10 & 5.952 \\
  5 & 1     &       &  -21.65 & 7.138e10 & 3.061 \\
    & & & & & \\
 15 & 0.5   & 0.109 &  -19.83 & 1.329e10 & 8.225 \\
 10 & 0.5   &       &  -20.21 & 1.890e10 & 5.786 \\
  5 & 0.5   &       &  -20.92 & 3.621e10 & 3.020 \\
    & & & & & \\
 15 & 0.1   & 0.022 &  -18.19 & 2.951e9  & 7.463 \\
 10 & 0.1   &       &  -18.56 & 4.119e9  & 5.347 \\
  5 & 0.1   &       &  -19.24 & 7.741e9  & 2.845 \\
    & & & & & \\
 15 & 0.05  & 0.011 &  -17.51 & 1.566e9  & 7.034 \\
 10 & 0.05  &       &  -17.86 & 2.170e9  & 5.077 \\
  5 & 0.05  &       &  -18.54 & 4.055e9  & 2.717 \\
    & & & & &  \\
\hline
\hline
\end{tabular}
\end{center}
\label{tab8}
\normalsize
\end{table}

%%%%%%%%%%%%Figure 14

%%%%%%%%%%%%%%%Table 9 (cumulative M/L_B)
\begin{table}
\vskip 0.3 cm
\begin{center}
\vskip 0.2 cm
\caption{The cumulative mass to blue-luminosity ratio $log(M/L_{B})_{\odot}$ 
at the age of 15 Gyr and as a function of the galacto-centric distance. 
$M_{L}(r_{j/2})$ in the mass in units of $10^{12}\times M_{\odot}$ contained 
in the sphere of radius $r_{j/2}$. The magnitudes and colours are the 
integrated values within the same sphere. The radii $r_{j/2}$ correspond to 
the central core, $0.6 R_{L,e}$, $R_{L,e}$ and $2 R_{L,e}$.}
\vskip 0.25 cm
\scriptsize
\begin{tabular} {r c c c c c c}
\hline
\hline
    &      &        &      &        &       &      \\
$M_{L,T,12}$ & $ r_{j/2}$ & $M_V$ & (B-V) & $M_B$ & $M_{L}(r_{j/2})$ & $M/L_B$ \\
\hline
    &      &        &      &        &       &       \\
3   & 0.06 & -21.07 & 1.00 & -20.07 & 0.123 & 7.448 \\
    & 0.58 & -22.60 & 1.00 & -21.64 & 0.632 & 8.976 \\
    & 1.04 & -22-69 & 0.99 & -21.70 & 0.656 & 8.856 \\
    & 1.48 & -22.70 & 0.99 & -21.70 & 0.660 & 8.803 \\
    &      &        &      &        &       &       \\
0.1 & 0.06 & -17.31 & 1.01 & -16.30 & 0.004 & 8.466\\
    & 0.58 & -18.96 & 0.93 & -18.03 & 0.021 & 8.141 \\
    & 1.04 & -19.08 & 0.91 & -18.17 & 0.022 & 7.511 \\
    & 1.48 & -19.10 & 0.91 & -18.19 & 0.022 & 7.462 \\
\hline
\hline
\end{tabular}
\end{center}
\label{tab9}
\normalsize
\end{table}

%********************************
\subsection{The UV excess}

All studied elliptical galaxies have detectable UV flux short-ward of about
$2000 \AA$ (Burstein et al. 1988) with large variation from galaxy to galaxy.
The intensity of the UV emission is measured by the colour (1550--V). Our
galaxy models are  compared in the plane  (1550--V) versus  $M_{V}$ to the
sample of galaxies  by Burstein et al. (1988), see  also Bender et al.
(1992, 1993).

\begin{figure}
%\picplace{9cm}
%\psfig{file=f_15_mv.ps,height=9.0truecm,width=8.5truecm}
\psfig{file=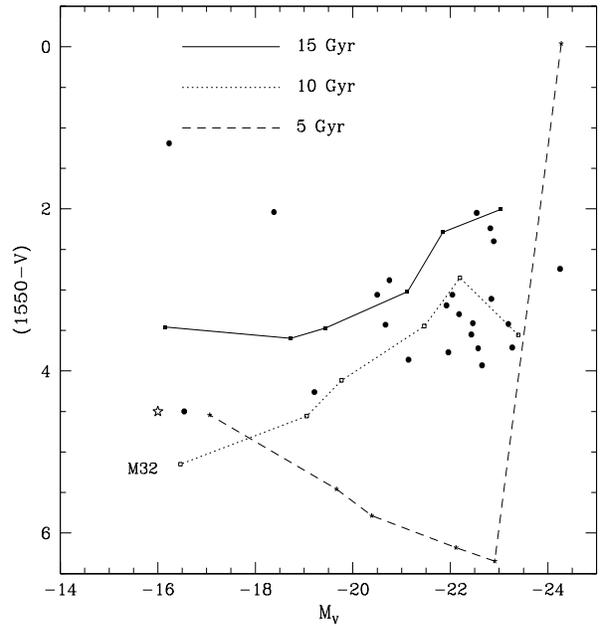,height=9.0truecm,width=8.5truecm}
\caption{The (1550--V) versus $M_V$ relation: the full dots are the data by 
Burstein et al. (1988), the lines show the theoretical predictions for the 
values of the age: 15 ({\em solid line}), 10 ({\em dotted line}), and 5 
({\em dashed line}) Gyr} 
\label{f_15_mv}
\end{figure}

The 1550 fluxes by Burstein et al. (1988) are derived from IUE data, which
refer to the region of a galaxy within  $14^{``}$ aperture. Assigning to the
galaxies of  Burstein et al. (1988) the distances calculated by Davies et
al. (1987), the IUE aperture roughly corresponds to a radius of  $\sim
1.234$ Kpc ($H_{0} = 50~ {\rm km/sec/Mpc }$ ). In contrast, the $V$ magnitudes 
refer to the whole galaxy and therefore a different kind of correction to the 
theoretical data is required (see the case of the CMR above). Having done that,
we derive the (1550-V) colours of our model galaxies  and compare it to the
observational data. This is shown in the  (1550--V) versus $M_{V}$ plane 
of Fig.~\ref{f_15_mv} for three different values of the age (15, 10, and 5 Gyr)
as indicated.

%%%%%%%%%%%%%%Figure 15

%*********************************
\subsection{Surface brightness}

One of the key assumptions of our models was the adoption of the Young 
(1976) density profile for the luminous material and the implicit use of the 
condition that the models should asymptotically reproduce the $R^{1/4}$ law.
Do the models match this constraint?

To answer the above question, we need to calculate the surface brightness
of our models as a function of the age. The method is as follows. Let for each 
shell $F_{\Delta \lambda}(j/2)$ and $M_{\Delta \lambda}(j/2)$ indicate the 
total flux emitted in the pass-band $\Delta \lambda$ and the corresponding 
magnitude, respectively,  

\begin{equation}
F_{\Delta \lambda}(j/2) = 10^{-0.4 M_{\Delta \lambda}(j/2)} 
\end{equation}

\noindent
where the flux $F_{\lambda}(j/2)$ is  $\rm erg/s/cm^{2}/str/\AA$. The flux per 
unit volume of each shell is 

\begin{equation}
\Omega_{F_{\Delta \lambda}}(j/2) = \frac{F_{\lambda}(j/2)}{\Delta V_{j/2}}
\end{equation}

\noindent
where $\Delta V_{j/2}$ is the volume in $\rm kpc^3$ of the $j$-th shell and
$\Omega_{F_{\Delta\lambda}}(j/2)$ is in $\rm erg/s/cm^{2}/str/\AA/kpc^{3}$.

Projecting the spherical shells onto a plane perpendicular to the line
of sight and passing through the center, we can define the elemental volume

\begin{equation}
dV = \frac{(r_{j+1}-r_{j})}{2} \times (r_{j+1}+r_{j}) \times d\theta
\times dl 
\end{equation}

\noindent
where $r_{j+1}$ and $r_{j}$ are the outer and inner radius of each shell,
$dl$ is the elemental length along the line of sight, and $\theta$ is the
angle between a given reference line passing through the center and drawn
on the above plane and any radial direction from the same center on the same
plane. The angle $\theta$ varies in the interval $0 < \theta < \pi$. Elementary
geometrical considerations set for the coordinate $l$ the range of variation
$0 < l < \sqrt{R^{2}-r_{j}^{2}}$, where $R$ is the external radius of the last
shell. With the aid of this the flux emerging from the $j$-th shell corrected
for the contribution from all overlaying layers is 

\begin{displaymath}
F_{tot, \lambda}(j/2) = 2 \times \int_{0}^{\pi} d\theta
\end{displaymath}
\begin{equation}
\times \int_{0}^{\sqrt{r_{tot}^{2}-r_{j}^{2}}} (r_{j+1}-r_{j}) \times
(r_{j+1}+r_{j}) \Omega_{F_{\lambda}}(l) dl
\end{equation}

\noindent
with obvious meaning of $\Omega_{F_{\Delta \lambda}}(l)$.

Known the total flux emerging from each shell (since this flux has been 
derived from absolute magnitudes it corresponds to a source located at the
distance of  $10pc$), we derive the apparent magnitude and finally the
surface brightness. To this aim, we fix an arbitrary distance $d$ and scale the
flux $F_{tot,\lambda}(j)$ of the ratio

\begin{displaymath}
  {1 \over 4\pi} \times { (10pc)^{2}  \over   (d)^{2} }
\end{displaymath}

\noindent
thus obtaining the flux emitted per unit solid angle by a source located at 
the distance $d$.
 
Finally, the surface brightness is given by

\begin{equation}
\mu_{\lambda}(j) = -2.5 \log{\left( \frac{F_{tot, \lambda}(j)
(10pc)^{2}}{d^{2} 4\pi} \frac{1}{\Sigma}\right)}
\end{equation}

\noindent
where $\Sigma$ is the apparent projected surface of the galaxy up to the 
$j$-th shell as it would appear at the distance $d$. Given the external radius 
$r_{j+1}$ of the shell, the corresponding angular surface up to that position 
is 

\begin{displaymath}
\pi(\Theta(arcsec))^{2} ~~~{\rm and} ~~~ \Theta = 206264.8 \frac{r_{j+1}}{d}
\end{displaymath} 

As expected the surface brightness $\mu_{\lambda}(j)$ does not depend on the 
arbitrary distance $d$ introduced to calculate the apparent flux.

\begin{figure}
%\picplace{9cm}
\psfig{file=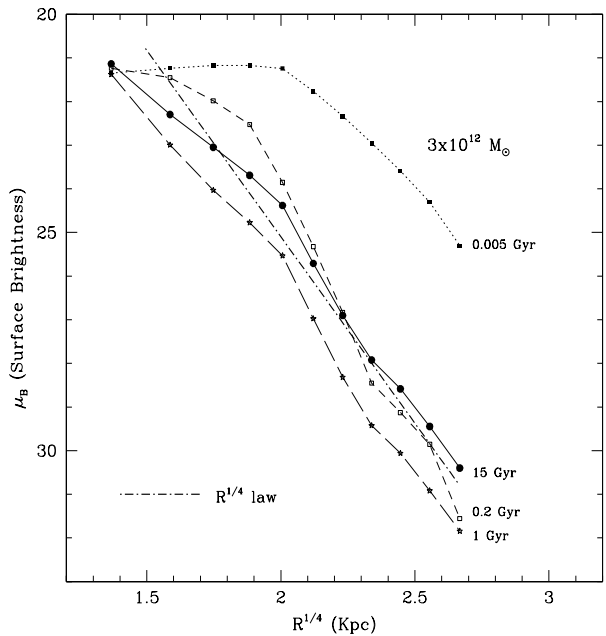,height=9.0truecm,width=8.5truecm}
\caption{The theoretical surface brightness profile for a model with
3$M_{L,T,12}$ at different ages as indicated. The dot-dashed line is the
$R^{1/4}$ law. Note how the models match the $R^{1/4}$ law when star
formation throughout the galaxy is completed}
\label{f_su_bri}
\end{figure}

The surface brightness obtained from the above procedure is shown in 
Fig.~\ref{f_su_bri} as a function of $(r/R_{e})^{1/4}$ for the case of the 
3$M_{L,T,12}$ galaxy at the age of 15 Gyr. For the purpose of comparison, we 
also plot the reference $R^{1/4}$ law in arbitrary units (heavy solid line).
{\it Despite the crudeness of our modelling the structure and evolution of elliptical 
galaxies, this fundamental condition is verified}.

%%%%%%%%%%%%Figure 16

In the same diagram we also plot the surface brightness  at other ages, i.e.
0.005 Gyr (very early stage), 0.2 Gyr, and 1 Gyr. Since the mass-luminosity
ratio is no longer constant in time and in space at these early epochs, larger
departures from the $R^{1/4}$ law are expected and noticed. Remarkably at
increasing galaxy age the luminosity profile gets closer and closer to the
$R^{1/4}$ law simply reflecting the fact that a constant mass to luminosity
ratio across the galaxy is gradually built up. Therefore we expect strong
departures  from the $R^{1/4}$ law over the period of time in which the front
of star formation activity recedes from the periphery toward the center.

%**********************************************************************
\subsection{Gradients in broad-band colours and line strength indices}

In this section we compare theoretical models and observational data in 
relation to the gradients in broad-band colours and line strength indices 
across individual galaxies. The discussion is limited to (B-R) and (1550-V), 
and \Hbeta, $\rm Mg_2$, and \MFe. 
To this aim we prefer to adopt a unique albeit 
small set of data, i.e. the five galaxies studied by Carollo \& Danziger 
(1994a), for the sake of internal homogeneity. The basic data for the galaxies 
in question are summarized in Table~10. Finally, we examine the distribution
of the galaxies in the \Hbeta\ versus [MgFe] plane and compare them with 
the predictions of theoretical models. The analysis is limited to the galaxies 
of the Gonz\'ales (1993) sample.  

%%%%%%%%%%%%%%%Table 10 Data of Carollo and Danziger galaxies
\begin{table*}
\vskip 0.3 cm
\begin{center}
\vskip 0.2 cm
\caption{Basic data for the Carollo \& Danziger (1994a) galaxies. The various 
quantities are: V the recession velocity; $R_{e}$" the effective radius in 
arcsec; $\sigma_0$ the central velocity dispersion; $D$ the distance in 
$10^4$ kpc; $R_{e}$ the effective radius in kpc; M the estimated mass in 
$10^{12} M_{\odot}$; $m_B$ and $M_B$ the apparent and absolute B-magnitudes, 
respectively; $L_B/L_{\odot}$ the blue luminosity in solar units; 
$(M/L_B)_{\odot}$ the blue mass to luminosity ratio.}
%\scriptsize
\begin{tabular*}{155mm}{l c c  c c c  c c c  c c}
\hline
\hline
& & & & & & & & & & \\
Name & V    & $R_{eff}$'' & $\sigma_0$ & $D$ & $R_{eff}$ & M & $m_B$ & $M_B$ & $L_B/L_{\odot}$ & $(M/L_B)_{\odot}$ \\
& & & & & & & & & & \\
     & km/s & & km/s & $10^4$kpc & kpc & $10^{12} M_{\odot}$ & & & & \\
& & & & & & & & & & \\
\hline
& & & & & & & & & & \\
NGC439 & 5679 & 45 & --- &11.36& 24.78&  --- & 12.38&--22.89 &2.24(11)& --- \\ 
NGC2434& 1388 & 24 & 205 &2.77 & 3.23 & 0.084& 12.50&--19.72 &1.20(10)& 7.00\\
NGC3706& 3046 & 27 & 281 &6.09 & 7.97 & 0.39 & 11.87&--22.05 &1.63(11)& 3.78\\
NGC6407& 4625 & 33 & --- &9.25 & 14.80& ---  & 12.88&--21.95 &9.38(10)& --- \\
NGC7192& 2761 & 28 & 185 &5.52 & 7.50 &0.16  & 12.21&--21.50 &6.20(10)& 2.56\\ 
& & & & & & & & & & \\
\hline
\hline
\end{tabular*}
\end{center}
\label{tab10}
%\normalsize
\end{table*}

%&&&&&&&&&&&&&&&&&&&&&&&&&&&&&&&&&&&&&&&&&&&&&
\subsubsection{Gradients in broad-band colours}

\begin{figure}
%\picplace{9cm}
\psfig{file=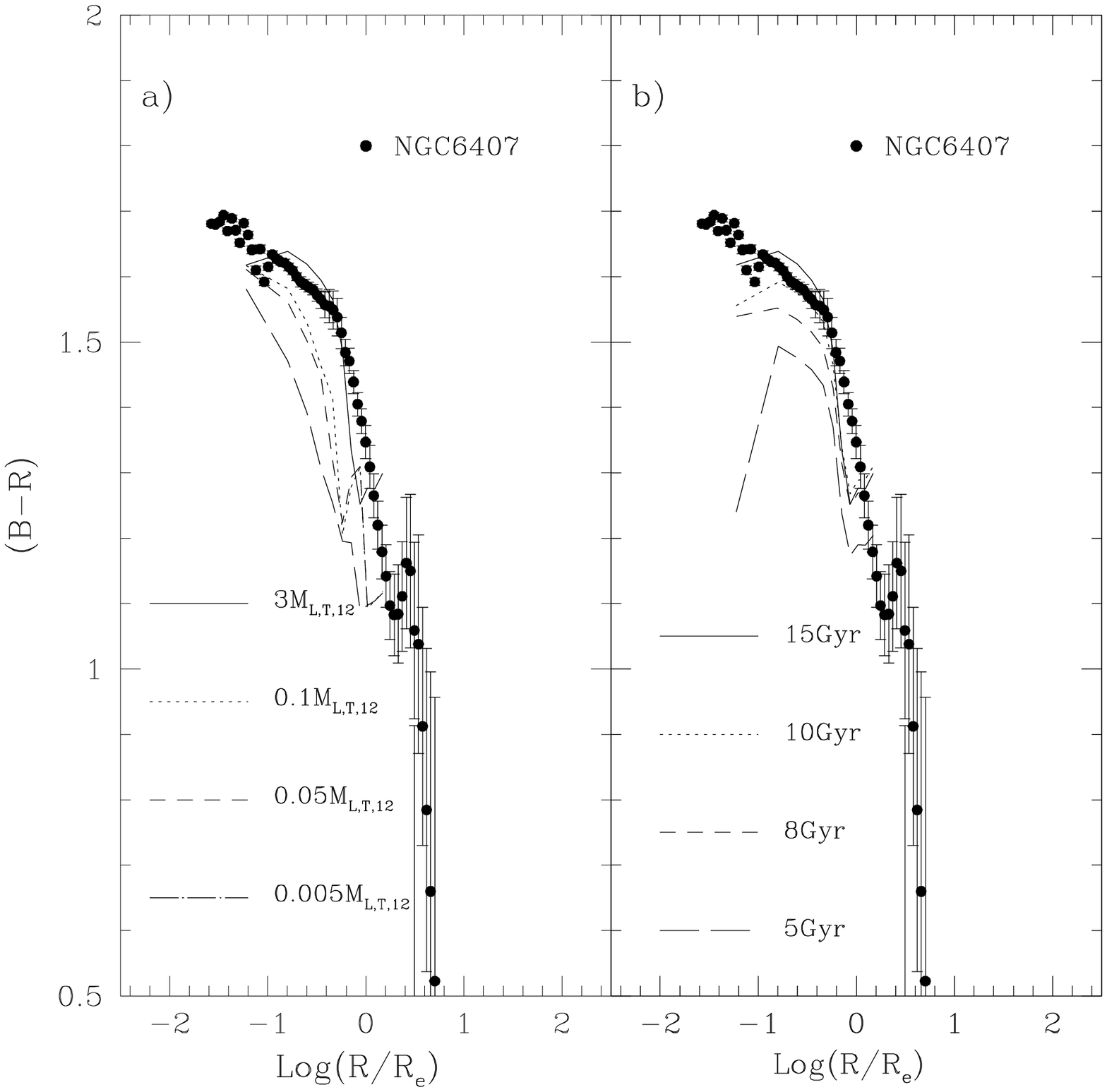,height=9.0truecm,width=8.5truecm}
\caption{The gradient in the (B--R) colour across the galaxy NGC~6407 (full 
dots). The data are from Carollo \& Danziger (1994a). Panel (a) shows the 
colour gradient for models of the same age (15 Gyr) and different $M_{L,T,12}$ 
as indicated. Panel (b) displays the colour gradient for the model with 
3$M_{L,T,12}$ and different ages: 15 ({\em solid line}), 10 ({\em dotted line}), 
and 5 ({\em dashed line}) Gyr}
\label{br_6407}
\end{figure}

\begin{figure}
%\picplace{9cm}
\psfig{file=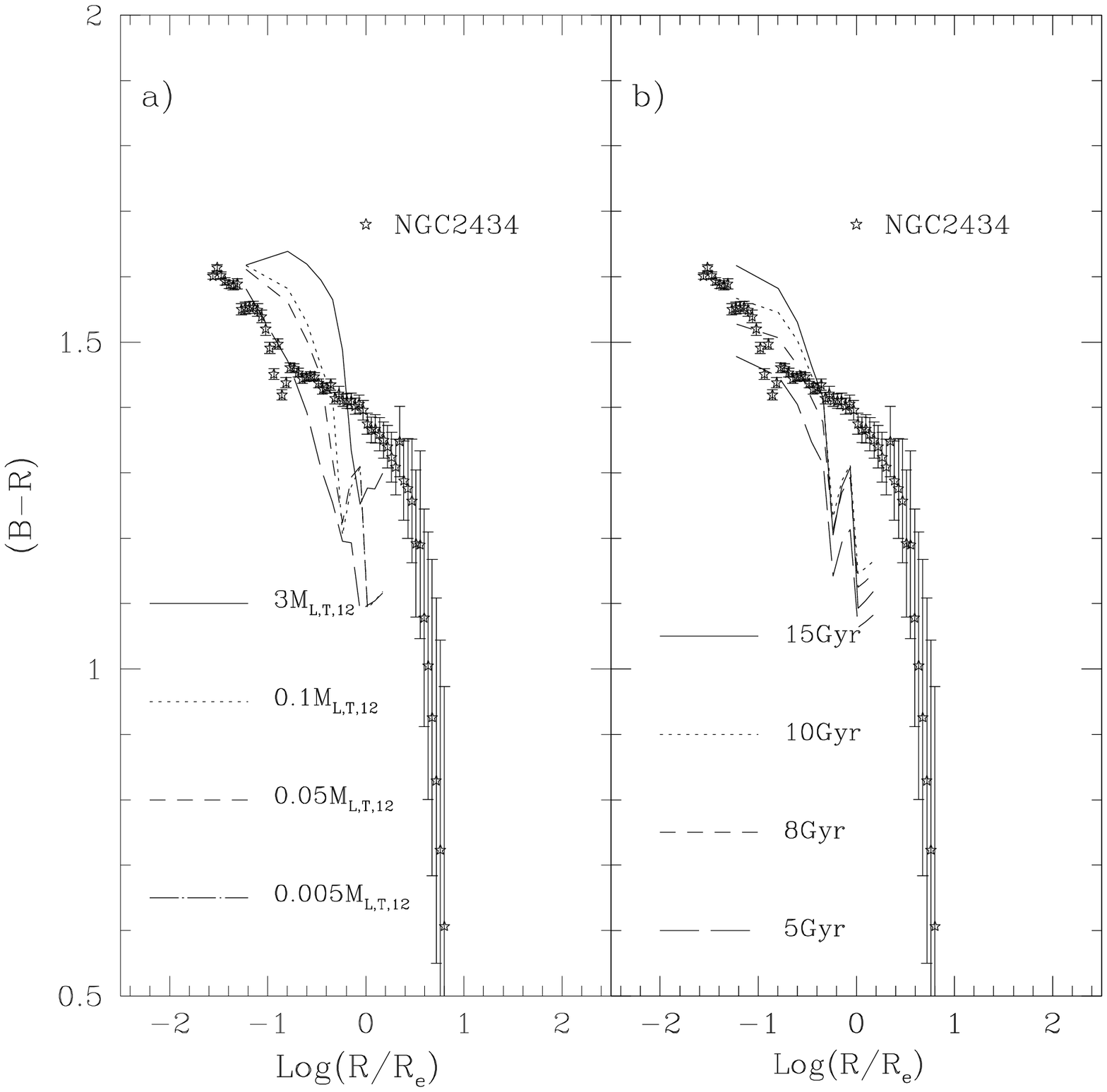,height=9.0truecm,width=8.5truecm}
\caption{The gradient in the (B--R) colour across the galaxy NGC~2434 (full
dots). The data are from Carollo \& Danziger (1994a). Panel (a) shows the
colour gradient for models of the same age (15 Gyr) and different $M_{L,T,12}$
as indicated. Panel (b) displays the colour gradient for the model with
0.1$M_{L,T,12}$ and different ages: 15 ({\em solid line}), 10 ({\em dotted
line}), and 5 ({\em dashed line}) Gyr}
\label{br_2434}
\end{figure}

To compare theoretical and observational colours we consider two galaxies of 
different luminosity and mass in turn, namely the high luminosity galaxy 
NGC~6407 shown in Fig.~\ref{br_6407} and the somewhat fainter object NGC~2434 
shown in Fig.~\ref{br_2434}. The colours are plotted as a function of the 
galacto-centric distance in units of effective radius. In Figs.~\ref{br_6407} 
and \ref{br_2434}, panels (a) show the data and the theoretical results for 
models of the same age (15 Gyr) and different mass, whereas panels (b) show the
same but for models of given mass and different age.

In order to choose the model galaxy best matching the observational data we 
have made use of the $(M/L_B)_{\odot}$ ratio and selected the models whose 
$(M/L_B)_{\odot}$ is comparable to the observational one. We find that 
NGC~6407 and NGC~2434 well correspond to the 3 and 0.1$M_{L,T,12}$ models, 
respectively. It is worth recalling that the two models have different 
properties: specifically star formation in the core stops at 5.12 and 1.45 Gyr 
and the mean metallicity is $Z_{mean}=0.036$ and 0.031 in the 3 and 
0.1$M_{L,T,12}$ models respectively. 

In the case of NGC~6407, it seems that the observational gradient is compatible
with that of the old age model (15 Gyr). In the case of NGC~2434 the situation
is less clear. First, there seems to be a systematic offset along the x-axis
perhaps caused by an uncertainty in the distance. We estimate that a 5\%
shorter distance would yield a better agreement. Second, the slope of the
colour gradient in the central regions is first flatter and then steeper than
indicated by old ages curves. Perhaps, the hint arises for a mean  age of the
stellar content in the region $-1.5 \leq \log{R/r_{e}} \leq 1$ younger by
several Gyr than in the outer regions. Recurrent or much prolonged episodes
of star formation activity in the central regions would lead to bluer colours.
The very central core require a slightly different explanation, because if
younger ages are invoked, they should be accompanied by significant metal
enrichment in order to get a red colour over there, which indeed is as red as
that of the old age case. The other galaxies of the sample show similar
problems. This more complex history of star formation cannot be described by
the present models, because they follow the classical SN-driven wind scheme
according to which star formation is monolithic and of shorter duration at
decreasing galaxy mass.

%&&&&&&&&&&&&&&&&&&&&&&&&&&&&&&&&&&&&&&&
\subsubsection{Gradients in (1550-V)}

\begin{figure}
%\picplace{9cm}
\psfig{file=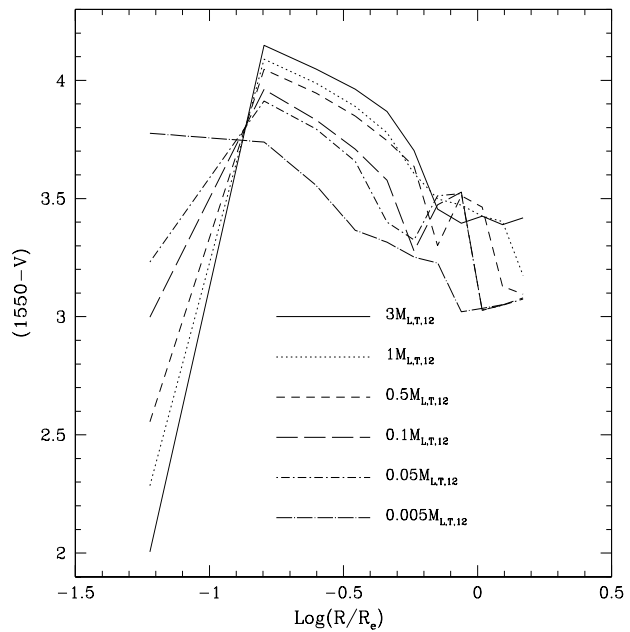,height=9.0truecm,width=8.5truecm}
\caption{The predicted gradient in the (1550--V) colour for the models with 
different $M_{L,T,12}$ as indicated}
\label{f_15_gra}
\end{figure}

Another interesting  gradient to look at is the one in the  (1550-V) colour,
 whose 
theoretical expectation is shown in Fig.~\ref{f_15_gra}. The easiest way to 
understand the behaviour of the (1550-V) colour across the model galaxies is 
by means of SSP's with different metallicity, cf. for instance Fig.~5 in 
TCBF96. Our models have the following basic features: (i) the metallicity 
(both mean and maximum) increases toward the center; (ii) the relatively early 
galactic winds but for the very central region (cf. the entries of Table 4 or 
Fig.~\ref{f_w_age}) secure that at the present age most of the galaxy is made 
by old stars with little age difference as we move inside (the oldest stars are
in the outermost regions), whereas the central region may contain stars over a 
much wider age range. However, even in this case the bulk population is 
relatively old because of the time dependence of the star formation rate. 
Therefore, for the whole galaxy but the center, to a first approximation we can
assume that all stars are nearly coeval but get more and more metal-rich going 
toward the center. In such a case the effect of an increasing metallicity is 
that the colour (1550-V) gets larger and larger. In the central core we have 
the combined effect of a higher metallicity and the presence of a younger 
stellar component. If the metallicity is $Z\leq 0.05$, the presence of younger 
ages would increase (1550-V) even further (cf. Fig.~5 in TCBF96). However, this
trend is destroyed by the presence of even small traces of high metallicity
stars (say $Z\geq 0.05$). In such  a case, AGB-manqu\'e and H-HB stars older
than about 5.6 Gyr (with the SSP's in usage) that are powerful, long-lived
sources of UV radiation reverse the trend in (1550-V) colour, which gets
``blue'' again. A detailed discussion of this effect can be found in Bressan
et al. (1994) and TCBF96 to whom the reader should refer. Observational data
on gradients in (1550-V) are not yet available to our knowledge.

%%%%%%%%%%%%Figure 17

%%%%%%%%%%%%Figure 18

%%%%%%%%%%%%%Figure 19

%&&&&&&&&&&&&&&&&&&&&&&&&&&&&&&&&&&&&&&&&&&&&&&&&&&&&&&
\subsubsection{Gradients in line strength indices}

Adopting the method described in Bressan et al. (1996), we have calculated
the temporal  and spatial evolution of the line strength indices in our model
galaxies. The definition of the line strength indices strictly follows Worthey 
(1992) and Worthey et al. (1994). In particular we made use of their fitting 
functions, in which there is no dependence on the possible enhancement in 
$\alpha$ elements with respect to iron expressed by [$\alpha$/Fe] with the 
usual meaning of the notation. Complete tabulations of the indices are
available from the authors upon request.

%%%%%%%%%%%%Figure 20
\begin{figure}
%\picplace{9cm}
\psfig{file=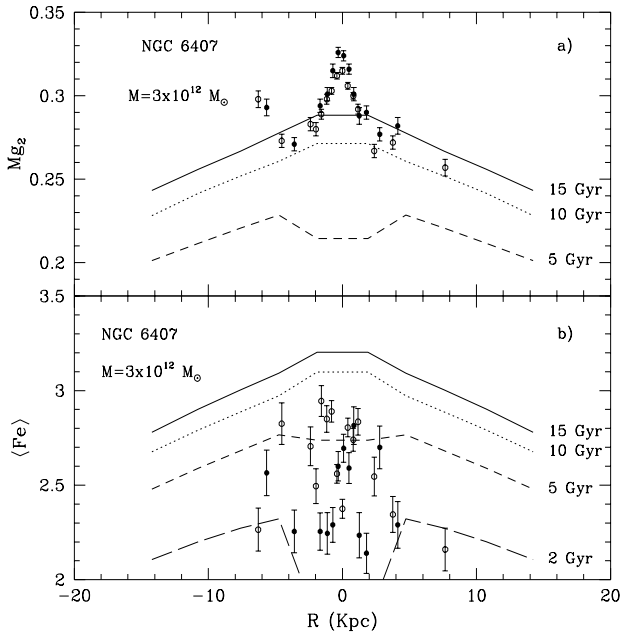,height=9.0truecm,width=8.5truecm}
\caption{The line indices \Mg2 and $\rm \langle Fe \rangle$ as function of the 
galacto-centric distance, {\em panel a)} and {\em panel b)} respectively, as 
measured by Carollo \& Danziger (1994a) along the major ({\em full dots}) and 
minor axis ({\em empty dots}) of NGC~6407. Superposed are the theoretical 
gradients at three different values of the age as indicated}
\label{f_mf_640}
\end{figure}

%%%%%%%%%%%%%Figure 21
\begin{figure}
%\picplace{9cm}
\psfig{file=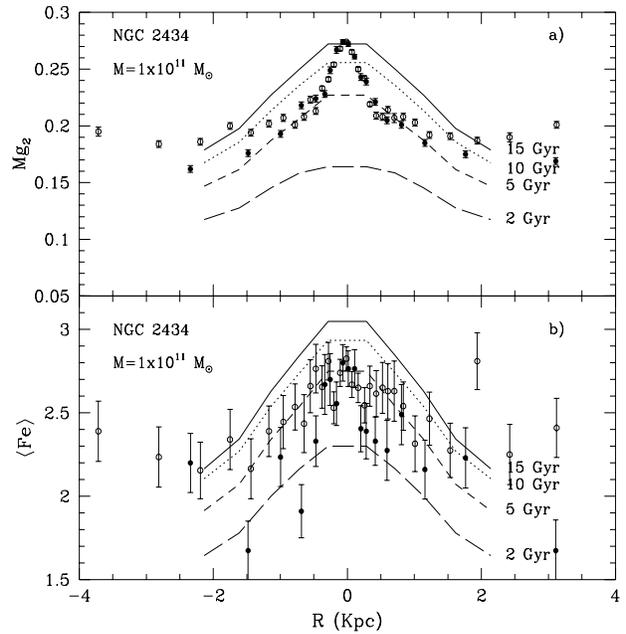,height=9.0truecm,width=8.5truecm}
\caption{The line indices \Mg2 and $\rm \langle Fe \rangle$ as function of the 
galacto-centric distance, {\em panel a)} and {\em panel b)} respectively, as 
measured by Carollo \& Danziger (1994a) along the major ({\em full dots}) and 
minor axis ({\em empty dots}) of NGC~2434. Superposed are the theoretical 
gradients at three different values of the age as indicated}
\label{f_mf_243}
\end{figure}

We compare here the gradients in \Mg2 and $\rm \langle Fe \rangle$ 
observed in the 
galaxies NGC~6407 (Fig.~\ref{f_mf_640}) and NGC~2434 (Fig.~\ref{f_mf_243}), 
with those predicted by the same models used in the analysis of the (B-R) 
colours. In both diagrams, the top panel is for \Mg2, the bottom
panel for $\rm \langle Fe \rangle$. The filled circles indicate the indices
measured along the major axis, while the open circles show the same but along
the minor axis. The theoretical gradients are displayed for several values of
the age as indicated.

It is soon evident that these models fail in reproducing the gradients in the 
\Mg2\ and $\rm \langle Fe \rangle$. The situation is nearly the same for both 
galaxies. Surprisingly, the disagreement is stronger for NGC~6407 for which 
the gradient in (B-R) was reproduced. No obvious causes of the failure can be 
found on the basis of the present calculations.

%&&&&&&&&&&&&&&&&&&&&&&&&&&&&&&&&&&&&&&&&&&
\subsubsection{The \Hbeta-\MgFe\ plane} 

\begin{figure}
%\picplace{9cm}
\psfig{file=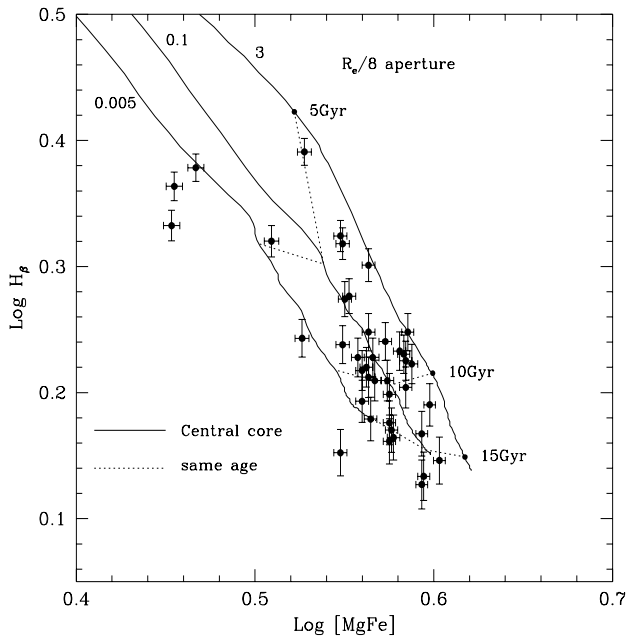,height=9.0truecm,width=8.5truecm}
\caption{Evolution of central region of the model galaxies (solid lines)  in 
the \Hbeta-\MgFe\ plane. The dotted lines are the loci of constant age as 
indicated. The full dots are the Re/8-data of Gonz\'ales (1993) with their 
observational uncertainties}
\label{f_hb_mf}
\end{figure}

Despite the above failure we look at the evolutionary path of the central 
region of the models in the \Hbeta-\MgFe\ plane and compare it with the
Re/8-data from the Gonz\'ales (1993) sample of elliptical galaxies, which is 
indicative of the central properties of the galaxies. The comparison is shown 
in Fig.~\ref{f_hb_mf} limited to the models with total mass of 3, 0.1 and 
0.005$M_{L,T,12}$. Since the theoretical indices are calculated using 
calibrations that do not take into account the possible enhancement in 
$\alpha$-elements which in contrast is suspected to exist in elliptical 
galaxies from the analysis of the line strength indices \Mg2 and 
$\rm \langle Fe 
\rangle$ and their gradients, see for instance the recent reviews by Matteucci 
(1994, 1997) and the above discussion, to somehow cope with this marginal 
discrepancy in plotting the theoretical results we have applied the offset 
$\rm \Delta \log{[MgFe]} = 0.05$ (cf. also Bressan et al. 1996).

Remarkably and even more intriguing, data and theoretical results seem to
agree each other, in the sense that the general trend shown by the data is 
recovered by the models.

The long debated question posed by this diagram is whether or not galaxies
span a large range of ages. The point is made evident looking at line of
constant age drawn in Fig.~\ref{f_hb_mf}. It is worth recalling that along each
theoretical sequence the age increases from the top to the bottom of the 
diagram.

Before going further, we clarify that the observational uncertainty cannot be 
the cause of the observed spread, along the \Hbeta\ axis in particular. 
According to Gonz\'ales (1993) the uncertainty in \MgFe\ is very small, not 
exceeding $\Delta \MgFe=\pm 0.03$, whereas that in \Hbeta\ is larger, but also 
in this case not exceeding $\Delta \Hbeta=\pm 0.06$. 

%%%%%%%%%%%%%Figure 22

The studies by Bressan et al. (1996) and Greggio (1996) on the distribution 
of galaxies in the \Hbeta\ versus \MgFe\ plane with the aid of SSP's and/or
complete models or both have come to a number of interesting yet demanding 
conclusions:

\begin{itemize}
\item {Galaxies span a small range of mean metallicities (Bressan et al. 1996;
Greggio 1966).}
\littleskip

\item {The percentage of low metallicity stars is small. This can be achieved
either by infall (TCBF96) or prompt enrichment (Greggio 1996).}
\littleskip

\item {The distribution of galaxies in the \Hbeta\ versus \MgFe\ plane does not
agree with the expectation from the CMR if this latter is the locus of old,
nearly coeval galaxies (see the 15 Gyr locus in Fig.~\ref{f_hb_mf}). However,
as the CMR we refer to is for cluster galaxies, whereas the Gonz\'ales sample
includes both cluster and field objects, the discrepancy is not conclusive,
see Bressan et al. (1996) for more details.}
\littleskip

\item {Based on the difference between the $R_e$/8 and $R_e$/2-data (this
latter sampling a wider area of the galaxies, cf. Gonz\'ales 1993 for details)
Bressan et al. (1996) suggested that in most galaxies the nucleus was younger,
or more precisely star formation lasted longer,  and more metal-rich than the
external regions.}
\littleskip

\item {Finally, Bressan et al. (1996) also proposed  that the overall duration
of the star formation activity, at least in the central regions, ought to be
increase at decreasing galaxy mass.}
\end{itemize}

In the present models, while star formation in the nucleus lasts longer thus
leading to higher metallicities than  in the external regions, still the total
duration of the star forming activity is shorter at decreasing galaxy mass
(the SN-driven wind scheme with constant IMF). Therefore, the questions posed 
by the \Hbeta\ versus \MgFe\ plane cannot be answered by the models in 
question even if with their spatial gradients in star formation they provide 
better leverage than the classical one-zone models. 

%%%%%%%%%%%%%%%%%%%%%%%%%%%%%%%%%%%%
\section{Summary and Conclusions}

In this paper we have described a simple multi-zone model of spherical
galaxies in which spatial gradients in mass density and star formation
are taken into account. The model, which is an extension of the classical
one-zone infall model, aims at following the history of star formation and
chemical enrichment taking place in primeval gas falling into the potential 
well of dark matter. 

Given a certain spherical distribution of dark matter supposedly constant with 
time, the primeval gas is let flow in and  gradually build up the spherical
distribution of baryonic mass originally in form of gas and later in form of 
stars. Since there is no dynamics in our model, the action consists in 
supposing that the mass of each spherical shell gradually increase at a 
suitable rate so that the final radial distribution of baryonic mass matches 
the one inferred from observational data for real galaxies. In the adopted 
scheme each shell is supposed to evolve without exchanging material with the
surrounding shells, in a sort of one-zone approximation. The lack of radial 
motions of gas toward the center is clearly the major drawback of our model, 
which eventually finds its justification only in the quality and robustness
of the final results as compared to observational data.  

The time of the local increase in baryonic mass seeks to closely follow the 
gross features emerging from fully dynamical models of galaxy formation: in a 
typical structure the radial velocity first increases to a maximum and then
decrease to zero at decreasing galacto-centric distance. This implies that the 
rate of mass accretion is a function of the radial distance, being large both 
at the center and in the external regions and small in between, the minimum 
being reached at a certain typical radius. 

The local rate of star formation is assumed to follow the Schmidt law, i.e.
$\dot \rho(r,t)_s =\nu(r) \rho_g(r,t)^{\kappa}$ with $\kappa=1$, where the
proportionality factor $\nu(r)$ is a suitable  function  of the galacto-centric
distance. It ultimately stems from the competition between the local free-fall 
and collision time scale as far as the global mode of star formation is 
concerned. This makes the specific efficiency of star formation increase
outward.

It is worth recalling that owing to the way the problem has been formulated
there are no free parameters in the accretion time scale and efficiency 
of star formation. Both follow from assigning the mass of the galaxy.

All other physical ingredients of the model, e.g. the ratio of dark matter to
baryonic mass, the IMF, and  the nucleosynthesis prescriptions are standard. 
They can be changed without altering the main scenario emerging from this 
study. The mass of dark matter and its spatial distribution simply affect the
total gravitational potential, and the epoch of galactic wind in turn. The IMF 
is the classical Salpeter law with $x=2.35$ and $\zeta=0.5$. This latter 
parameter is the fraction of mass in the IMF above 1$M_{\odot}$, driving the 
nucleosynthesis enrichment and the mass-luminosity ratio. As already recalled, 
the value $\zeta=0.5$ has been chosen so that the mean mass to blue luminosity 
ratios for the elliptical galaxies of the Bender et al. (1992, 1993) sample are
matched. 

Care has been paid to check whether the baryonic component of the model 
satisfies its most demanding constraints: (i) constant mass-luminosity ratio 
with the radius; (ii) $R^{1/4}$ luminosity profile. Considering the fully 
independent scheme on which the building up of the luminous component and 
associated star formation are based, the successful matching of the above 
constraints perhaps hints that the model despite its crudeness is on the right 
track and can be safely utilized to predict the gross features of the spatial 
gradients in metallicity, partition function $N(Z)$, ages, broad-band colours, 
and line strength indices of spherical (elliptical as well) galaxies to be 
compared with observational data.   
Main results of this study are:

(1) Galactic winds occur later (more precisely star formation is halted 
because the local thermal content of the gas exceeds the gravitational 
potential) as we move inward, i.e. galaxies are viewed as an {\it outside-in 
process}, in such a way that the older populations are expected in the external
regions of the galaxy. Likewise, the star forming activity lasts longer at 
increasing galactic mass. In this context, the present models strictly conform 
to the classical supernova driven galactic wind scenario.
 
(2) As a result, the external regions contain stars that are richer in
$\alpha$-elements than the inner ones, and the more massive models are on 
the average less enhanced in $\alpha$-elements than the low mass ones. 
Also in this respect, the present models conform  the expectation from the 
classical supernova driven galactic wind scenario. This is not specific to our 
model but common to all chemical models with monolithic star formation and
constant IMF. Line strength indices such as \Mg2\ and 
$\rm \langle Fe \rangle$ seem
to require the opposite trend: more $\alpha$-enhancement toward the center in 
more massive galaxies. 

(3) The color-magnitude relation, and the mean mass to blue luminosity ratios
can be easily matched. The CMR is best explained by old, nearly coeval galaxies
of different mean metallicity (increasing with the galaxy luminosity and hence
mass). However, concerning the $M/L_{B}$ ratio, like all other models of this
type (i.e. with constant IMF) the increase by approximately a factor of 3 
to 5 passing
from low to high mass elliptical galaxies (the so-called tilt of the FP)
cannot be reproduced without relaxing the notion of coevality.

(4) The observational \Hbeta\ and \MgFe\ for the central regions of elliptical
galaxies are reproduced  by the models, which however are still not able 
to answer the question posed by \Hbeta-\MgFe\ plane, i.e. if and why elliptical
galaxies  may have undergone star formation over long periods of time, some 
in the far past other in more recent epochs.

(5) The models are marginally able to reproduce gradients in broad band
colors but have serious difficulties as far as the gradients in the line
strength gradients are concerned. The nature  of the disagreement 
is not easy to 
understand because all the models possess well behaved gradients in star 
formation (lasting
longer toward the center) and mean and maximum metallicity (increasing toward
the center).
It seems as if in addition to the gross scheme we have considered, 
real galaxies had a  more complicated history of star formation to which 
the line strength indices are more sensitive than the broad band colors
or the integrated properties. Reconstructing the kind of  star 
formation that  satisfies all the constraints imposed by observational
data of  real galaxies is a cumbersome affair beyond the capability of
the present model.
Owing to its importance, the whole problem is addressed in a 
companion paper (Tantalo et al. 1998), in which the effects of different 
calibrations for the indices as a function of effective 
temperature, gravity, metal 
content, and degree of enhancement in $\alpha$-elements of the constituent 
stars, and different histories of star formation and chemical enrichment of 
the models have been thoroughly scrutinized. No details of this investigation 
are given here for the sake of brevity.

In summary, despite its gross simplification of the physics leading to the
formation of a galaxy,   the model we have presented successfully reproduces
some of the properties of real galaxies but fails in others, the gradients
in line strength indices in particular. 
However, considering the global performance of the model, 
 we are confident that it can be safely utilized to investigate the 
properties of elliptical   under
different assumptions concerning the IMF, the past history of star 
formation (monolithic in the early past  or recurrent in several episodes),
and the SN-driven galactic wind mechanism.

\smallskip

\acknowledgements{C.C. is pleased to acknowledge the kind hospitality and the 
highly stimulating environment provided  by ESO in Garching where this paper 
was written up during sabbatical leave from the Astronomy Department of the 
Padua University. This study has been funded by the Italian Ministry of
University, Scientific Research and Technology (MURST), the Italian Space
Agency (ASI), and the European Community TMR grant \# ERBFMRX-CT96-0086.}

\end{document}